\documentclass[a4paper,11pt]{article}
\pdfoutput=1 
\usepackage{jcappub} 
\usepackage{lineno}

\usepackage{tikz}
\usepackage{color,soul}
\usepackage[english]{babel}
\usepackage{blindtext}
\usepackage{longtable}
\usepackage{multirow}
\usepackage{comment}
\usepackage{float}
\usepackage{physics}
\usepackage{textcomp, gensymb}
\usepackage{subcaption} 
\usepackage{tabularx}
\usepackage{enumitem}

\newcommand{\oii}{[OII]}
\newcommand{\loii}{$L_{\rm [OII]}$}
\newcommand{\lgl}{$\log L_{\rm [OII]}$}
\newcommand{\lgms}{$\log M_*$}

\newcommand{\gfib}{$g_{\rm fib}$}
\newcommand{\dz}{$D_z^{grz}$}
\newcommand{\dstar}{$D_s^{grz}$}
\renewcommand{\arcsec}{$^{\prime\prime}$}
\newcommand{\Ngal}{$N_{\rm gal}$}

\newcommand{\hmpc}{$h^{-1}$Mpc}

\title{\boldmath DESI Data Release 2 ELGs: Property-dependent subsamples, imaging systematics, and clustering}

\author[a]{T.~Hagen} 
\emailAdd{tyler.hagen@utah.edu}

\author[a]{K.~S.~Dawson} 
\emailAdd{kdawson@astro.utah.edu}

\author[a]{Z.~Zheng} 
\emailAdd{zhengzheng@astro.utah.edu}

\author[b]{J.~Aguilar}
\author[c]{S.~Ahlen} 
\author[d,e]{D.~Bianchi} 
\author[f]{D.~Brooks}
\author[b]{T.~Claybaugh}
\author[g]{A.~de la Macorra} 
\author[h,i]{B.~Dey} 
\author[b,j]{S.~Ferraro} 
\author[k,l]{J.~E.~Forero-Romero} 
\author[m]{S.~{Gontcho~A~Gontcho}} 
\author[n]{G.~Gutierrez}
\author[b]{J.~Guy} 
\author[o]{C.~Hahn} 
\author[p]{M.~Ishak} 
\author[q]{R.~Joyce} 
\author[q]{S.~Juneau} 
\author[b]{A.~Kremin} 
\author[f]{O.~Lahav} 
\author[r]{C.~Lamman} 
\author[b]{M.~Landriau} 
\author[s]{L.~Le~Guillou} 
\author[t,u]{M.~Manera} 
\author[q]{A.~Meisner} 
\author[v,u]{R.~Miquel}
\author[w]{J.~Moustakas} 
\author[x]{A.~D.~Myers}
\author[y]{S.~Nadathur} 
\author[i]{J.~ A.~Newman} 
\author[z,aa]{G.~Niz} 
\author[ab,ac,ad]{W.~J.~Percival} 
\author[b,ae,j]{C.~Poppett}
\author[af]{F.~Prada} 
\author[ag]{I.~P\'erez-R\`afols} 
\author[ah,ai,r]{A.~J.~Ross} 
\author[aj]{G.~Rossi}
\author[ak]{S.~Saito} 
\author[al]{E.~Sanchez} 
\author[b]{D.~Schlegel}
\author[b]{J.~Silber} 
\author[am]{G.~Tarl\'{e}} 
\author[q]{B.~A.~Weaver}
\author[an]{H.~Zou} 

\affiliation[a]{Department of Physics and Astronomy, University of Utah, 270 South 1400 East, Salt Lake City, UT 84112, USA}
\affiliation[b]{Lawrence Berkeley National Laboratory, 1 Cyclotron Road, Berkeley, CA 94720, USA}
\affiliation[c]{Department of Physics, Boston University, 590 Commonwealth Avenue, Boston, MA 02215 USA}
\affiliation[d]{Dipartimento di Fisica ``Aldo Pontremoli'', Universit\`a degli Studi di Milano, Via Celoria 16, I-20133 Milano, Italy}
\affiliation[e]{INAF-Osservatorio Astronomico di Brera, Via Brera 28, 20122 Milano, Italy}
\affiliation[f]{Department of Physics \& Astronomy, University College London, Gower Street, London, WC1E 6BT, UK}
\affiliation[g]{Instituto de F\'{\i}sica, Universidad Nacional Aut\'{o}noma de M\'{e}xico,  Circuito de la Investigaci\'{o}n Cient\'{\i}fica, Ciudad Universitaria, Cd. de M\'{e}xico  C.~P.~04510,  M\'{e}xico}
\affiliation[h]{Department of Astronomy \& Astrophysics, University of Toronto, Toronto, ON M5S 3H4, Canada}
\affiliation[i]{Department of Physics \& Astronomy and Pittsburgh Particle Physics, Astrophysics, and Cosmology Center (PITT PACC), University of Pittsburgh, 3941 O'Hara Street, Pittsburgh, PA 15260, USA}
\affiliation[j]{University of California, Berkeley, 110 Sproul Hall \#5800 Berkeley, CA 94720, USA}
\affiliation[k]{Departamento de F\'isica, Universidad de los Andes, Cra. 1 No. 18A-10, Edificio Ip, CP 111711, Bogot\'a, Colombia}
\affiliation[l]{Observatorio Astron\'omico, Universidad de los Andes, Cra. 1 No. 18A-10, Edificio H, CP 111711 Bogot\'a, Colombia}
\affiliation[m]{University of Virginia, Department of Astronomy, Charlottesville, VA 22904, USA}
\affiliation[n]{Fermi National Accelerator Laboratory, PO Box 500, Batavia, IL 60510, USA}
\affiliation[o]{Department of Astronomy, University of Texas at Austin, 2515 Speedway, TX 78712, USA}
\affiliation[p]{Department of Physics, The University of Texas at Dallas, 800 W. Campbell Rd., Richardson, TX 75080, USA}
\affiliation[q]{NSF NOIRLab, 950 N. Cherry Ave., Tucson, AZ 85719, USA}
\affiliation[r]{The Ohio State University, Columbus, 43210 OH, USA}
\affiliation[s]{Sorbonne Universit\'{e}, CNRS/IN2P3, Laboratoire de Physique Nucl\'{e}aire et de Hautes Energies (LPNHE), FR-75005 Paris, France}
\affiliation[t]{Departament de F\'{i}sica, Serra H\'{u}nter, Universitat Aut\`{o}noma de Barcelona, 08193 Bellaterra (Barcelona), Spain}
\affiliation[u]{Institut de F\'{i}sica dâ€™Altes Energies (IFAE), The Barcelona Institute of Science and Technology, Edifici Cn, Campus UAB, 08193, Bellaterra (Barcelona), Spain}
\affiliation[v]{Instituci\'{o} Catalana de Recerca i Estudis Avan\c{c}ats, Passeig de Llu\'{\i}s Companys, 23, 08010 Barcelona, Spain}
\affiliation[w]{Department of Physics and Astronomy, Siena University, 515 Loudon Road, Loudonville, NY 12211, USA}
\affiliation[x]{Department of Physics \& Astronomy, University  of Wyoming, 1000 E. University, Dept.~3905, Laramie, WY 82071, USA}
\affiliation[y]{Institute of Cosmology and Gravitation, University of Portsmouth, Dennis Sciama Building, Portsmouth, PO1 3FX, UK}
\affiliation[z]{Departamento de F\'{\i}sica, DCI-Campus Le\'{o}n, Universidad de Guanajuato, Loma del Bosque 103, Le\'{o}n, Guanajuato C.~P.~37150, M\'{e}xico}
\affiliation[aa]{Instituto Avanzado de Cosmolog\'{\i}a A.~C., San Marcos 11 - Atenas 202. Magdalena Contreras. Ciudad de M\'{e}xico C.~P.~10720, M\'{e}xico}
\affiliation[ab]{Department of Physics and Astronomy, University of Waterloo, 200 University Ave W, Waterloo, ON N2L 3G1, Canada}
\affiliation[ac]{Perimeter Institute for Theoretical Physics, 31 Caroline St. North, Waterloo, ON N2L 2Y5, Canada}
\affiliation[ad]{Waterloo Centre for Astrophysics, University of Waterloo, 200 University Ave W, Waterloo, ON N2L 3G1, Canada}
\affiliation[ae]{Space Sciences Laboratory, University of California, Berkeley, 7 Gauss Way, Berkeley, CA  94720, USA}
\affiliation[af]{Instituto de Astrof\'{i}sica de Andaluc\'{i}a (CSIC), Glorieta de la Astronom\'{i}a, s/n, E-18008 Granada, Spain}
\affiliation[ag]{Departament de F\'isica, EEBE, Universitat Polit\`ecnica de Catalunya, c/Eduard Maristany 10, 08930 Barcelona, Spain}
\affiliation[ah]{Center for Cosmology and AstroParticle Physics, The Ohio State University, 191 West Woodruff Avenue, Columbus, OH 43210, USA}
\affiliation[ai]{Department of Astronomy, The Ohio State University, 4055 McPherson Laboratory, 140 W 18th Avenue, Columbus, OH 43210, USA}
\affiliation[aj]{Department of Physics and Astronomy, Sejong University, 209 Neungdong-ro, Gwangjin-gu, Seoul 05006, Republic of Korea}
\affiliation[ak]{Institute for Multi-messenger Astrophysics and Cosmology, Department of Physics, Missouri University of Science and Technology, 1315 N Pine St, Rolla, MO 65409 USA}
\affiliation[al]{CIEMAT, Avenida Complutense 40, E-28040 Madrid, Spain}
\affiliation[am]{University of Michigan, 500 S. State Street, Ann Arbor, MI 48109, USA}
\affiliation[an]{National Astronomical Observatories, Chinese Academy of Sciences, A20 Datun Road, Chaoyang District, Beijing, 100101, P.~R.~China}

\abstract{
Using emission-line galaxies (ELGs) from the Dark Energy Spectroscopic Instrument (DESI) Data Release 2, we evaluate a property-dependent correction to imaging systematics.
We derive systematic weights following the same linear regression method used for other DESI tracers, but do so separately on ELG subsamples to provide a physically-informed alternative to the fiducial, neural-network-based approach.
In doing so, we show that the deeper imaging in the Dark Energy Survey (DES) footprint leads to a higher overall number density but a lack of targets with extreme $g-r$ and $r-z$ colors.
ELGs in the DES region also show a distinct redshift distribution when subsampled by position in the $g-r$ vs. $r-z$ plane.
To address these effects, we implement a separate treatment of the DES footprint within the DESI catalog production pipeline, which is generally well-motivated and, in some cases, imperative for accurate clustering measurements.
With DES treated separately, we find that property-dependent systematic weights further mitigate spurious clustering signal in $\sim$10\% of subsamples, while the fiducial scheme remains optimal for the full sample.
}

\keywords{galaxy clustering, redshift surveys} 

\begin{document}
\maketitle
\flushbottom

\section{Introduction}\label{sec:intro}
Modern galaxy redshift surveys such as the Baryon Oscillation Spectroscopic Survey (BOSS; \cite{dawson13}), extended Baryon Oscillation Spectroscopic Survey (eBOSS; \cite{dawson16}), and Dark Energy Spectroscopic Instrument (DESI; \cite{desi16_science, desi16_instrument}) survey map the three-dimensional distribution of galaxies in an effort to measure galaxy clustering, probe the large-scale structure (LSS) of the Universe, and constrain cosmological parameters.
However, survey effects can induce fluctuations in the projected galaxy density field and must be mitigated to ensure clustering measurements represent the true cosmological signal. 
One such effect is broadly known to be associated with variations in image quality, in which the photometry used to identify spectroscopic targets suffers from an angular dependence on the foreground Milky Way (e.g., Galactic extinction and stars) and survey itself (e.g., depth and seeing).
If left uncorrected, these inhomogeneities can lead to apparent over- or underdensities in the galaxy field that correlate with imaging properties (e.g., see Figure~6 in \cite{raichoor23}).
Thus, to accurately probe the fundamental physics of the Universe, artificial clustering signal induced by imaging systematics must be separated from the cosmological signal.

Many strategies have been employed to mitigate spurious signal from imaging data and its impact on galaxy clustering.
Popular methods make use of angular maps of each imaging property to either derive per-galaxy weights before the clustering measurement or directly apply a correction to the clustering measurement post facto.
Such methods are summarized in \cite{weaverdyck21} and include: linear (or non-linear; see, e.g., \cite{hernandez-monteagudo14, nicola16}) regression in which projected galaxy density is modeled as a linear function of imaging properties and inverted to compute weights (e.g., \cite{ross11, laurent17, ross17, elvin-poole18}); template subtraction in which the cross correlation between imaging properties (decomposed into angular harmonics) is subtracted from the clustering measurement (e.g., \cite{ross11, ho12}); and mode projection, which can be considered as a non-decomposed version of template subtraction (e.g., \cite{rybicki92, leistedt13, elsner17}).
More recent mitigation strategies employ forward models that inject sources into the imaging data (e.g., \cite{burleigh18, kong20, kong25}) and machine learning algorithms such as neural networks (e.g., \citep{rezaie20, sysnet, rosado-marin25}) and random forests (e.g., \citep{chaussidon22}).
With any strategy, assessing the impact on clustering measurements and areas for improved accuracy will remain critical as the size and precision of spectroscopic surveys increase.

Conducting the largest galaxy redshift survey to date, DESI is primarily designed to measure two distinct signals in LSS: the baryon acoustic oscillation (BAO; \cite{desi_dr1iii_galbao, desi_dr1iv_lyabao, desi_dr1vi_baocosmo, desi_dr2i_lyabao, desi_dr2ii_baocosmo}) feature and redshift space distortions (RSD; \cite{desi_dr1v_fs, desi_dr1vii_fscosmo}), which probe cosmic expansion and growth of structure, respectively.
While the BAO feature is relatively robust against imaging systematics (e.g., \cite{ross12, ross17, andrade25}), two-point clustering over a range of scales---such as those relevant to RSD and primordial non-Gaussianity \citep{bermejo-climent25, chaussidon25, chiarenza25}---remains sensitive to spurious fluctuations in imaging. 
To mitigate the effects, the DESI Collaboration develops so-called systematic weights to be applied during the clustering measurement.
In the latest DESI BAO analysis \citep{desi_dr2ii_baocosmo}, the systematic weights for all but one DESI target class are derived using multilinear regression on the galaxy density against multiple relevant imaging properties \cite{desi_dr1ii_samples}.
However, as the faintest galaxy tracer in the DESI survey, emission-line galaxies (ELGs) possess fluxes close to the imaging depth of the target selection survey, making these galaxies particularly sensitive to imaging systematics \citep{raichoor23}.
To derive systematic weights for such a complex sample, DESI instead employs \textsc{SYSNet}, a fully connected Feed Forward Neural Network \citep{sysnet, rosado-marin25}.
While such treatment captures non-linear correlations between galaxy density and imaging properties, the increased freedom of machine learning also risks overcorrection and removal of true cosmological signal \citep{rezaie24}.
Further, \cite{desi_dr1ii_samples} note that the clustering of DESI ELGs from the first year of observations shows spurious signal on scales beyond that of the BAO, which they attribute to residual imaging systematic errors not captured by the current weighting. 
Thus, exploring alternative methods with freedom beyond linear regression is well-motivated.

In this work, we explore a novel extension to the DESI ELG systematic weight derivation: customization according to galaxy property.
To do so, we rederive the systematic weights of the DESI Data Release 2 (DR2) ELGs using the same linear regression employed for all other DESI tracers.
However, rather than treating all ELGs as a single sample (as is done in the fiducial approach), we perform the regression separately on distinct subsamples defined by galaxy properties.
Doing so provides an alternative mechanism for flexibility beyond standard linear regression, reduces the risk of overfitting from machine learning, and can be compared with the fiducial, neural-network-based approach.
Our method is also physically motivated; the extent to which an ELG is impacted by imaging systematic errors may depend on the physical properties of the galaxy itself.
For example, \cite{hagen25} report scale-dependent bias in the clustering of ELGs from the DESI Early Data Release \citep{desi_edr}, the amplitude of which is largest for low- and intermediate-stellar-mass subsamples; such behavior may indicate the need for customized systematic weights.
Although we focus on DESI, property-dependent treatment of imaging systematics is also being explored for other surveys, such as the Vera Rubin Observatory's Legacy Survey of Space and Time \citep{kong26}.
Throughout this work, we consider five galaxy properties that describe both the physical and observed state of ELGs, with which we define ELG subsamples, compare rederived weights to their fiducial counterparts, and inspect consequences in the resultant clustering.

This paper is structured as follows.
In Section~\ref{sec:data}, we describe the DESI data, ELGs, and galaxy properties of interest.
In Section~\ref{sec:weights}, we summarize the weighting scheme applied to the clustering measurement, including the derivation of property-dependent systematic weights.
We present our analysis of property-dependent ELG subsamples in the fixed redshift range of $0.8<z<1.1$ in Section~\ref{sec:analysis}.
We also inspect two additional considerations: variable behavior of ELGs across photometric source and the impacts on the clustering of the full ELG sample.
Conclusions are given in Section~\ref{sec:conclusions}.

In this work, we adopt a flat Lambda cold dark matter ($\Lambda$CDM) cosmology consistent with \cite{planck20}, using $\Omega_{m,0}=0.310$, $H_0 = 100h$ km s$^{-1}$ Mpc$^{-1}$, and $h=0.677$. 

\section{DESI data}\label{sec:data}

\subsection{DESI survey and Data Release 2} \label{subsec:dr2}
DESI is attached to the 4-meter Mayall telescope at Kitt Peak National Observatory and is used to conduct an eight-year survey in which the spectra of $\sim$60 million galaxies and quasars will be measured across $\sim$17,000 deg$^2$ of the sky \citep{desi_sv}.
To efficiently do so, the DESI focal plane hosts robotically-controlled fibers distributed across ten spectrographs, enabling the spectra of 5000 targets in a $\sim$8-$\deg^2$ field-of-view to be measured simultaneously \citep{silber23, miller24, poppett24}.
Each spectrograph includes three wavelength channels jointly sensitive from 3600--9800~\text{\AA} with a spectral resolution of $\lambda / \Delta \lambda \sim 2000$--5000 \citep{desi22}.

Optical and mid-infrared photometry are used to select targets for spectroscopic observation.
Optical imaging in $g$, $r$, and $z$ is provided by the DESI Legacy Imaging Surveys \citep{dey19}, which combine data from the Dark Energy Camera (DECam; \cite{flaugher15}) Legacy Survey (DECaLS), Beijing Arizona Sky Survey (BASS; \cite{zou17}), and Mayall $z$-band Legacy Survey (MzLS; \cite{dey16}).
The southern region of the footprint also makes use of deeper imaging from the Dark Energy Survey (DES; \cite{des05}).
Due to the distinct sites and instruments, separate consideration is often given to the ``North'' (BASS and MzLS) and ``South'' (DECaLS or DES) regions when concerning the photometry.
These data are coupled with mid-infrared $W1$, $W2$, $W3$, and $W4$ imaging from the Wide-Field Infrared Survey Explorer (WISE; \cite{wright10}).
According to this joint photometry, DESI defines four classes of extragalactic targets: the Bright Galaxy Survey (BGS; \cite{hahn23}), Luminous Red Galaxies (LRGs; \cite{zhou23}), ELGs \citep{raichoor23}, and quasars \citep{chaussidon23}.

The DESI survey is strategically divided into so-called tiles, which define a unique sky location and fiber-to-target assignments \citep{schlafly23}.
The spectroscopic data are processed according to \cite{guy23}.
Each spectrum is fit by the \textsc{Redrock}\footnote{\url{https://github.com/desihub/redrock}} software, which classifies the object and estimates the redshift using templates derived from principal component analysis of synthetic (for galaxies; \citep{anand24}) or real (for quasars; \citep{brodzeller23}) spectra.

To measure clustering statistics and constrain cosmological parameters, LSS ``clustering'' catalogs are created following \cite{ross25}.
As part of the catalog production, several weights are derived to optimize the signal-to-noise ratio in the clustering measurements and correct for unphysical artifacts induced by, e.g., survey incompleteness and inhomogeneities due to imaging.
Each LSS catalog is also accompanied by ``random'' catalogs that match the data's footprint and line-of-sight number density, $n(z)$, but are otherwise randomly distributed; such catalogs are necessary because the clustering measurement is made with respect to a random distribution.


DESI has measured the redshifts of over 30 million galaxies and quasars to complete the DR2 sample, which is internally designated as ``Loa.''
These data have already led to measurement of the BAO scale in the spatial distribution of galaxies and Lyman-$\alpha$ forest, which are described in \cite{desi_dr2ii_baocosmo} and \cite{desi_dr2i_lyabao}, respectively.
The DR2 data is accompanied by several value-added catalogs (VACs), including those created from the \textsc{FastSpecFit} code\footnote{\url{https://github.com/desihub/fastspecfit}}.
This code---the results of which we use in this work---estimates galaxy properties via modeling of the dust-corrected stellar continua and emission lines in the DESI spectra and photometry \citep{moustakas23}.

\subsection{DESI ELGs} \label{subsec:dr2elgs}
ELGs are a population of star-forming galaxies whose abundance and distinct spectral features enable studies of the $z \gtrsim 1$ Universe.
In particular, emission of the \oii{} doublet at rest-frame wavelengths of 3726 and 3729~\AA{} provides an unambiguous signature to estimate redshifts \citep{raichoor23}.
Within DESI, this sample is the largest of any target class and is defined to trace LSS primarily between $0.8 < z < 1.6$ \citep{desi_dr1ii_samples}.
Beyond their use as a cosmological tracer, DESI ELGs themselves have been the subject of many studies, spanning contexts such as their large-scale environment \citep{gonzalez-perez20}, relation to an underlying star-forming population \citep{lan24, yuan25}, and galaxy-halo connection (e.g., \cite{rocher23, gao23, gao24, hagen25, ortega-martinez25}).

The DESI ELG target selection relies on imaging data that are sensitive to the observing conditions during the DESI Legacy Surveys.
Fluctuations in these imaging conditions can alter the observed target density, and if left uncorrected, contaminate the measured clustering with non-cosmological signal.
Moreover, because the ELG fluxes approach the depth limits of the imaging survey, this sample is most vulnerable to such variations in image quality.
For this reason, we focus on ELGs.

As described in \cite{raichoor23}, DESI ELG targets are largely defined by cuts in (dust-corrected) $g-r$, $r-z$, and $g$-band fiber magnitude, \gfib{}.
To balance redshift coverage across target classes, DESI assigns relatively low priority to ELG targets during fiber assignment, denoting the primary sample as $\tt{ELG\_LOP}$.
An extension to redder colors is also defined to include ELGs with redshifts as low as 0.6 but with very low priority ($\tt{ELG\_VLO}$), as LRGs are the primary tracer at such redshifts.
The selection in $g-r$ vs. $r-z$ distinguishes the two ELG samples.
In both cases, $r-z>0.15$ and $g-r < 0.5 \left( r-z \right) + 0.1$ are required, which separate the sample from high-redshift ELGs (for which the [OII] doublet is outside DESI's wavelength coverage) and the stellar locus, respectively.
Specific to low-priority ELGs ($\tt{ELG\_LOP}$), $g-r < -1.2 \left( r-z \right) + 1.3$ is required to optimize the fraction of [OII] emitters with redshifts between 1.1 and 1.6.
Finally, a requirement in fiber magnitude, $g_{\rm fib} < 24.1$, is imposed for all ELG targets to ensure the sample favors \oii{} emitters and reaches the desired target density.

From the LSS catalogs (version $\tt{v2.1}$), we exclusively study the $\tt{ELG\_LOP}$ sample to ensure homogeneity, as the $\tt{ELG\_VLO}$ sample occupies a different region of the $g-r$ vs. $r-z$ plane \citep{raichoor23}.
Following \cite{desi_dr1ii_samples}, we further reject objects that also qualify for DESI's quasar target selection \citep{chaussidon23}; both choices are consistent with the main DESI cosmological analysis and result in the purest version of the sample, $\tt{ELG\_LOPnotqso}$, which we refer to as ELGs hereafter.

\subsection{ELG properties for subsampling} \label{subsec:properties}
We identify five galaxy properties according to which ELG subsamples will be defined.
Among these properties are stellar mass $\left( M_* \right)$ and \oii{} luminosity $\left( L_{\rm \left[ OII \right]} \right)$, which describe the evolutionary state of these star-forming galaxies.
Derived estimates of stellar mass and dust-corrected \oii{} flux ($F_{\rm [OII]}$) are taken from \textsc{FastSpecFit} \citep{moustakas23}.
We calculate \oii{} luminosities following \cite{hagen25}, which uses $r$-band photometry from the DESI Legacy Surveys \citep{dey19} to apply a customized aperture correction that accounts for the angular size of each ELG: 
\begin{equation} \label{eqn:loii}
    L_{\rm [OII]} = 4 \pi D_L^2(z) \left( \frac{F_{\rm [OII]}}{1.266} \cdot \frac{F_{\mathrm{total,}~r}}{F_{\mathrm{fiber,}~r}} \right).
\end{equation}
Here, $D_L(z)$ is the luminosity distance, $F_{\mathrm{total,}~r}$ is the total $r$-band flux, and $F_{\mathrm{fiber,}~r}$ is the predicted $r$-band flux within the DESI fiber in 1\arcsec{} seeing.
The factor of 1.266 backs out the original aperture correction, which assumes a point source and is inaccurate for extended ELGs.

The remaining three properties are motivated by the DESI ELG target selection and make use of the DESI Legacy Survey photometry \citep{dey19}.
We consider \gfib{} because it traces the ELG brightness and receives a faint limit of 24.1 in the target selection.
Finally, we define two quantities that describe position in the $g-r$ vs. $r-z$ target selection plane.
The first quantity is the perpendicular distance from the boundary that optimizes the ELG redshift range, $g-r < -1.2 (r-z) + 1.3$:
\begin{equation}
    D_z^{grz} = -0.7682 \left( r-z \right) - 0.6402 \left( g-r \right) + 0.8322.
\end{equation}
Similarly, the second quantity is the perpendicular distance from the boundary that separates ELG targets from the stellar locus, $g-r < 0.5 (r-z) + 0.1$:
\begin{equation}
    D_s^{grz} = 0.4472 \left( r-z \right) - 0.8944 \left( g-r \right) + 0.0894.
\end{equation}

\begin{figure*}
\begin{center}
    \includegraphics[width=0.85\textwidth]{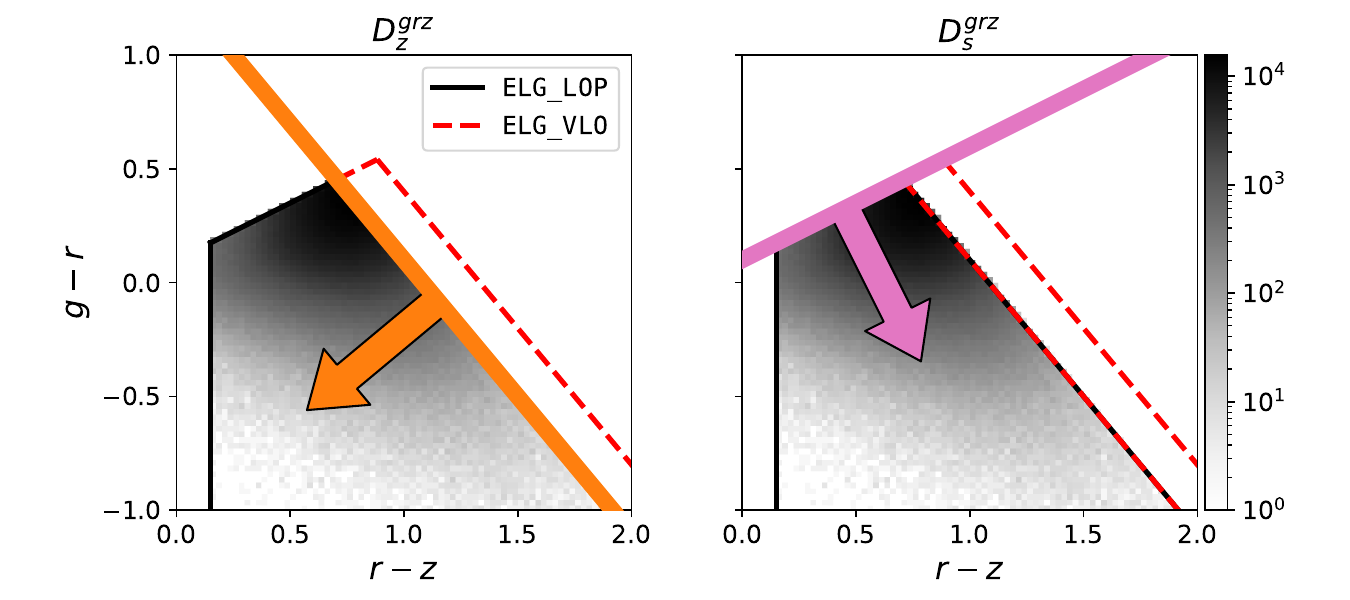}
    \caption{Illustration of \dz{} (left panel) and \dstar{} (right panel) in the $g-r$ vs. $r-z$ ELG target selection plane. Each solid colored line and arrow mark the zero-point and direction of increasing distance for the corresponding quantity. Shading indicates the number of $0.8<z<1.1$ ELGs ($\tt{ELG\_LOPnotqso}$) in two-dimensional bins and uses logarithmic scaling.}
    \label{fig:illustrate}
\end{center}
\end{figure*}

As illustrated in Figure~\ref{fig:illustrate}, \dz{} and \dstar{} increase as ELGs reside farther from the respective boundary, and a value of zero indicates colors that lie on the boundary. 
By using this color-color basis, we can directly examine the relationship between variations in imaging quality and contamination across the ELG target selection boundaries \citep{raichoor23}.

Throughout this work, we use the five properties described above---$M_*$, \loii{}, \gfib{}, \dz{}, and \dstar{}---to define subsamples, derive property-dependent systematic weights, and inspect the resultant clustering.
In Section~\ref{sec:analysis}, we do so at fixed redshift; we select a range consistent with the main DESI cosmological analysis, requiring $0.8<z<1.1$ (``ELG1'' in \cite{desi_dr1ii_samples}).
Table~\ref{tab:props} lists each property and corresponding 68\% and 95\% bounds for this sample.
Before presenting results, we first summarize the entire DESI weighting scheme---which is applied during the clustering measurement---in the following section.

\begin{table}
    \centering
    \begin{tabular}{|c|cccccc|}
    \hline
    Property & Description & $p_{2.5}$ & $p_{16}$ & $p_{50}$ & $p_{84}$ & $p_{97.5}$ \\
    \hline
    $\log M_*$ & Logarithmic stellar mass & 8.62 & 9.09 & 9.60 & 10.24 & 10.84 \\
    $\log L_{\rm [OII]}$ & Logarithmic [OII] luminosity & 41.47 & 41.71 & 41.93 & 42.17 & 42.43 \\
    \gfib{} & $g$-band fiber magnitude & 22.70 & 23.35 & 23.79 & 24.02 & 24.09 \\
    \dz{} & Distance from redshift boundary & 0.01 & 0.05 & 0.16 & 0.35 & 0.58 \\
    \dstar{} & Distance from stellar boundary & 0.01 & 0.08 & 0.21 & 0.41 & 0.73 \\
    \hline
    \end{tabular}
    \caption{The five galaxy properties considered in this work. In the right columns, $p_X$ is the $X$th percentile of each property for $0.8<z<1.1$ ELGs.}
    \label{tab:props}
\end{table}

\section{Summary of DESI weights} \label{sec:weights}
Although this work focuses on systematic weights, the DESI weighting scheme includes many additional components that mitigate survey effects and optimize the signal-to-noise ratio of the clustering measurement.
Complete descriptions of the full weighting scheme and ELG-specific systematic weights are provided in \cite{ross25, desi_dr1ii_samples} and \cite{rosado-marin25}, respectively.
We summarize the framework here.

In this work, we measure the two-dimensional, two-point correlation function (2PCF), which describes the probability of a galaxy pair being separated by given distances perpendicular ($r_p$) and parallel ($r_\pi$) to the line of sight.
We use the Landy-Szalay estimator \citep{landy93}, which counts galaxy pairs within and across the data and random catalogs,
\begin{equation} \label{eqn:landy}
    \xi (r_p,~r_\pi) = \frac{DD \left( r_p,~r_\pi \right) - 2DR \left( r_p,~r_\pi \right) + RR \left( r_p,~r_\pi \right)}{RR \left( r_p,~r_\pi \right)},
\end{equation}
where $DD \left( r_p,~r_\pi \right)$, $DR \left( r_p,~r_\pi \right)$, and $RR \left( r_p,~r_\pi \right)$ are the fraction of data-data, data-random, and random-random pairs with separations in a bin centered on $\left( r_p,~r_\pi \right)$, respectively.
The weights described below are applied when computing the pair counts: $DD$, $DR$, and $RR$.

\subsection{Systematic weights} \label{subsec:wsys}
Systematic weights mitigate fluctuations in the ELG density field caused by inhomogeneous imaging properties.
To do so, the projected number density of galaxies is generally modeled as a function of several imaging properties.
For DESI ELGs, ten properties are considered:
\begin{itemize}[label={}, leftmargin=0pt, itemsep=1ex]
    \item (i) stellar density ($\tt{STARDENS}$) constructed from Gaia Data Release 2 \citep{gaia18};
    \item (ii) column density of neutral Hydrogen ($\tt{HI}$) combined from the Effelsberg-Bonn HI Survey and Galactic All-Sky Survey \citep{hi4pi16};
    \item (iii--v) 5$\sigma$ detection depth---corrected for dust---in the DESI Legacy Survey DR9 $g$, $r$, and $z$ filters ($\tt{GALDEPTH\_\{GRZ\}}$) \citep{dey19};
    \item (vi--viii) full-width at half maximum of the PSF in the DESI Legacy Survey DR9 $g$, $r$, and $z$ filters ($\tt{PSFSIZE\_\{GRZ\}}$) \citep{dey19}; and
    \item (ix--x) the difference in Galactic extinction between the Schlegel-Finkbeiner-Davis (SFD) map \citep{schlegel98} and that estimated by \cite{zhou25} using DESI stars (which is estimated separately using two colors, $g-r$ and $r-z$, resulting in two differences, $\tt{EBV\_DIFF\_GR}$ and $\tt{EBV\_DIFF\_RZ}$, respectively).
\end{itemize}
Each property is mapped onto \textsc{Healpix} \citep{healpix05} pixels with $\textsc{Nside}=256$ resolution and \textsc{Nested} ordering (see Appendix~A of \cite{desi_dr1ii_samples} for details).
The number of spectroscopically confirmed galaxies in each pixel is calculated, after which the (pixelized) number density is modeled as a function of imaging properties.
Finally, the model is inverted to compute the corresponding weight, which is inherited by all galaxies in a given pixel.
Below, we summarize the two modeling strategies compared throughout this work; in both cases, the modeling is done separately for the North and South photometric regions due to their respective imaging coming from distinct instruments.

Unlike all other DESI tracers, the regression method for ELGs relies on \textsc{SYSNet}---a fully connected Feed Forward Neural Network \citep{sysnet}---and is not restricted to linear models.
Implemented in \cite{rosado-marin25}, each pixel is assigned the observed number of galaxies as a label and the values of each imaging map as features.
The model is trained, validated, and tested using five permutations of $k$-fold cross-validation with $k=5$.
This procedure is done on all ELGs as a single sample (within broad redshift bins), ultimately producing the systematic weights provided in the LSS catalogs ($\tt{WEIGHT\_SYS}$) and denoted as $w^{\rm all}_{\rm sys,~SYSNet}$ throughout this work.

In contrast, we rederive the systematic weights using the same linear regression as all other DESI tracers (and closely matching that used in eBOSS; \cite{bautista18, ross20}) but add flexibility by doing so separately for property-dependent subsamples.
Automated in the DESI LSS catalog production pipeline, this method divides the pixels into ten bins of each imaging property and performs a multilinear regression on the number density against each property simultaneously \citep{desi_dr1ii_samples, ross25}.
To ensure other survey effects do not drive the regression, all other individual-level weights ($w_{\rm FKP}w_{\rm tot}/w_{\rm sys}$ in Section~\ref{subsec:wother}) are applied during this procedure.
By separating the regression for each property-dependent subsample, we add physically motivated, additional freedom to the model while avoiding the risks of overfitting often attributed to machine learning.
We denote the resultant weights as $w^{\rm subsample}_{\rm sys,~linear}$ and compare to their fiducial, neural-network-based counterparts in Section~\ref{sec:analysis}.
Before doing so, we summarize the remaining components of the DESI weighting scheme.

\subsection{Additional weights}\label{subsec:wother}
In addition to systematic weights, a combination of individual, pairwise, and angular weights are used within DESI to ensure accurate measurement of galaxy clustering on all scales. \\

\noindent \textbf{Redshift Failure.} The removal of galaxies with unreliable redshift estimates induces spurious fluctuations in the observed density field which typically correlate with, e.g., effective observing time, number of exposures, focal plane position, and redshift.
To remove these fluctuations, each galaxy is weighted by a redshift failure weight, $w_{\rm zfail}$, which is derived by modeling the redshift success rate as a function of several spectroscopic observational quantities.
This derivation is detailed in \cite{yu25} and \cite{krolewski25} for ELGs and all other DESI tracers, respectively. \\

\noindent \textbf{Completeness.} Each galaxy is given a completeness weight to correct for unequal probability of fiber assignment during spectroscopic observation.
In DESI, completeness can be estimated by running 128 realizations of the fiber assignment algorithm, resulting in a series of bits indicating whether a galaxy received (bit value of 1) or did not receive (0) a fiber in each realization \citep{lasker25}.
An individual-inverse-probability (IIP) weight can then be calculated as one over the fraction of realizations in which a given galaxy received a fiber:
\begin{equation} \label{eqn:iip}
    w_{\rm comp,~IIP}^{m} = \frac{1 + 128}{1 + \texttt{popcount}\left( b^m \right)}.
\end{equation}
Here, $b^m$ is the bit series of galaxy $m$, \texttt{popcount}($b^m$) counts the number of realizations in which the galaxy received a fiber, and an offset of one is applied to the numerator and denominator to represent the survey observation itself.
The IIP completeness weight is useful for describing the overall completeness of a sample and sufficient for accurate clustering measurements if restricted to only large scales. \\

\noindent \textbf{Regional Random-to-Data Balance.} Each object in the random catalogs also requires the above weights ($w_{\rm sys}, w_{\rm zfail}, w_{\rm comp,~IIP}$) to ensure the weighted $n(z)$ follows that of the data.
Thus, each random object receives the redshift and above weights of a galaxy randomly sampled from the data.
This sampling is done separately in each photometric region which, for DESI ELGs, includes the North and South.
After this per-region sampling, the random catalogs receive an additional, oft-unnamed factor which sets the (weighted) random-to-data ratio equal between each photometric region.
The product of all other weights in the random catalog is multiplied by this extra factor, which we denote $w_{\rm reg}$. \\

Each galaxy receives two additional weights that are calculated separately for those in the North Galactic Cap (NGC) and South Galactic Cap (SGC), which lie above and below the Galactic plane, respectively. \\

\noindent \textbf{Feldman-Kaiser-Peacock.} First, standard ``Feldman-Kaiser-Peacock'' (FKP) weights, $w_{\rm FKP}$, are applied to optimize the signal-to-noise of two-point clustering measurements \citep{feldman94}.
Such weights are a function of number density, which is largely driven in DESI by the properties of the target sample and the completeness associated with the number of tiles on which each target appears ($N_{\rm tile}$).
Thus, the calculation of $w_{\rm FKP}$ accounts for $N_{\rm tile}$:
\begin{equation} \label{eqn:fkp}
    w_{\rm FKP} = \frac{1}{1 + n \left( z,~N_{\rm tile} \right) P_0}.
\end{equation}
Here, $P_0$ is a chosen power spectrum monopole amplitude and for DESI ELGs, is set to $P_0 \left( k=0.15~h^{-1}\rm{Mpc} \right) = 4000$ $\left( h^{-1}\rm{Mpc} \right)^{-3}$ to optimize measurement at the BAO scale \citep{desi_dr1ii_samples}. \\

\noindent \textbf{Average Completeness Division.} Second, because $N_{\rm tile}$ largely drives the completeness weights (e.g., ELGs with $N_{\rm tile}=6$ have a higher average completeness than those with $N_{\rm tile}=1$) and Eq.~(\ref{eqn:fkp}) is already a function of $N_{\rm tile}$, the average completeness per $N_{\rm tile}$ should be removed from each galaxy's total weight.
This factor, $1~/~\langle w_{\rm comp,~IIP} \rangle \left(N_{\rm tile} \right)$, also equates the \textit{weighted} number of galaxies to the \textit{observed} number of galaxies, further optimizing the clustering signal-to-noise. \\

Combining many of the components above yields the final $w_{\rm tot}$ defined in \cite{ross25, desi_dr1ii_samples}:
\begin{equation}
    w_{\rm tot} = \frac{w_{\rm zfail}w_{\rm sys}w_{\rm comp,~IIP}}{\langle w_{\rm comp,~IIP} \rangle \left( N_{\rm tile} \right)}.
\end{equation}
We note that $w_{\rm tot}$ of the randoms also includes the multiplication by $w_{\rm reg}$.
Despite being omitted in the above definition, the multiplication of $w_{\rm FKP}$ is also done for the data and randoms when performing the clustering measurement.

All weights described thus far are applicable to individual galaxies, but to accurately measure small-scale clustering, two additional components are needed: pairwise completeness weights and angular upweighting.
In practice, we follow the ``pairwise inverse probability with angular upweighting'' scheme in \cite{bianchi25} and Appendix C of \cite{desi_dr1ii_samples} to implement both components. \\

\noindent \textbf{Pairwise Completeness.} While IIP completeness weights apply to single galaxies, the assignment of a fiber to a given galaxy removes the possibility of other reachable galaxies receiving that same fiber in a single pass, yielding an underestimation of close pairs.
Therefore, to ensure unbiased small-scale clustering, this correction is best done using pairwise-inverse-probability (PIP) weights.
Unlike IIP weighting, each \textit{pair} is boosted according to one over the fraction of realizations in which \textit{both} galaxies were observed,
\begin{equation}
    w^{m,n}_{\rm comp,~PIP} = \frac{1+128}{1 + \texttt{popcount}\left( b^m~\&~b^n \right)},
\end{equation}
where \texttt{popcount}($b^m~\&~b^n$) counts the number of realization with \textit{joint} observation of galaxies $m$ and $n$ \citep{bianchi17}.
This formalism replaces the IIP weight when counting data-data and data-random pairs; random-random pairs receive the product of each object's IIP weight. \\

\noindent \textbf{Angular Upweight.} Some pairs possess zero probability of joint observation due to, e.g., physical limits of the fiber positioners.
Thus, angular upweighting is necessary to recover small-scale clustering signal that remains missing \citep{percival17, mohammad20}.
This correction leverages the full catalog of (pre-fibered) targets, the angular clustering of which is unaffected by fiber assignment.
By measuring the number of data-data and data-random pairs in the target catalogs (``parent'') as a function of angular separation $\theta$ and dividing by those of what receives a fiber (``fibered'')---i.e., $DD_{\rm parent}(\theta)~/~DD_{\rm fibered}(\theta)$ and $DR_{\rm parent}(\theta)~/~DR_{\rm fibered}(\theta)$---the true small-scale clustering can be recovered.
Random-random pairs do not require this correction because their angular positions are sampled from the target catalog \textit{before} fiber assignment.
These ratios are evaluated using 40 evenly spaced logarithmic bins of $\theta$ from $10^{-4}$--$10^{0.5}$ deg.
To properly assign weights in a given area, the average total and FKP weights for a given $N_{\rm tile}$ are computed from the LSS data catalog as $\langle w_{\rm tot}~/~w_{\rm comp,~IIP} \rangle \left( N_{\rm tile} \right)$ and $\langle w_{\rm FKP} \rangle \left( N_{\rm tile} \right)$, respectively, and applied to the target catalogs during the angular measurement \cite{desi_dr1ii_samples}.
In this work, we apply the same angular up-weighting to all ELG subsamples.
While a target catalog could be customized to match cuts in photometric properties (e.g., \gfib{}), the same cannot be done for estimated galaxy properties (e.g., \lgms{}) because some targets lack spectroscopy.
This choice has minimal consequence because the angular upweights deviate significantly from unity only on angular scales below $\sim$10$^{-3}$ deg, which corresponds to projected comoving scales of $r_p \sim 0.03$ \hmpc{} and $\sim$0.06 \hmpc{} at $z=0.8$ and 1.6, respectively; both scales are below what we consider in this work. \\

Table~\ref{tab:weights} summarizes the weights applied to pair counts when performing the two-point clustering measurement.
We emphasize that this entire scheme is standard in the DESI LSS catalog pipeline and yields the most accurate clustering measurement across all scales.

\begin{table}
    \centering
    \begin{tabular}{|ccccc|}
    \hline
    Weight & Description & $DD$ & $DR$ & $RR$ \\
    \hline
    $w_{\rm sys}$ & Imaging systematic & \checkmark & \checkmark & \checkmark \\
    $w_{\rm zfail}$ & Redshift failure & \checkmark & \checkmark & \checkmark \\
    $w_{\rm comp,~IIP}$ & Individual completeness & & \checkmark & \checkmark \\
    $w_{\rm reg}$ & Regional random-to-data balance & & \checkmark & \checkmark \\
    $w_{\rm FKP}$ & Feldman-Kaiser-Peacock & \checkmark & \checkmark & \checkmark \\
    $1~/~\langle w_{\rm comp,~IIP} \rangle \left( N_{\rm tile}\right)$ & $N_{\rm tile}$-averaged completeness & \checkmark & \checkmark & \checkmark \\
    $w_{\rm comp,~PIP}$ & Pairwise completeness & \checkmark & & \\
    $XX_{\rm parent}(\theta)~/~XX_{\rm fibered}(\theta)$ & Angular upweight & \checkmark & \checkmark & \\
    \hline
    \end{tabular}
    \caption{Summary of DESI weights applied during data-data ($DD$), data-random ($DR$), and random-random ($RR$) pair counting to measure the two-point clustering. We emphasize that the angular upweight---for which $XX$ represents $DD$ or $DR$---has additional subtlety; the calculation leverages the target catalog, which inherits $N_{\rm tile}$-averaged weights from the LSS data catalog. Refer to the text for details.}
    \label{tab:weights}
\end{table}

\section{Analysis of ELG subsamples} \label{sec:analysis}
We present our analysis of property-dependent ELG subsamples in the redshift range of \mbox{$0.8<z<1.1$}.
We first define subsamples according to each galaxy property, derive customized systematic weights, and compare to the fiducial, neural-network-based result.
The clustering under each weighting scheme is then compared.
Finally, we investigate subsamples defined by distance from the stellar locus (\dstar{}), which shows potential for isolating spurious signal, before presenting the implications on the clustering of the ELG sample as a whole. 

\subsection{Property-dependent subsamples, randoms, and weights} \label{subsec:weight_comp}
\begin{figure*}
\begin{center}
    \includegraphics[width=0.85\textwidth]{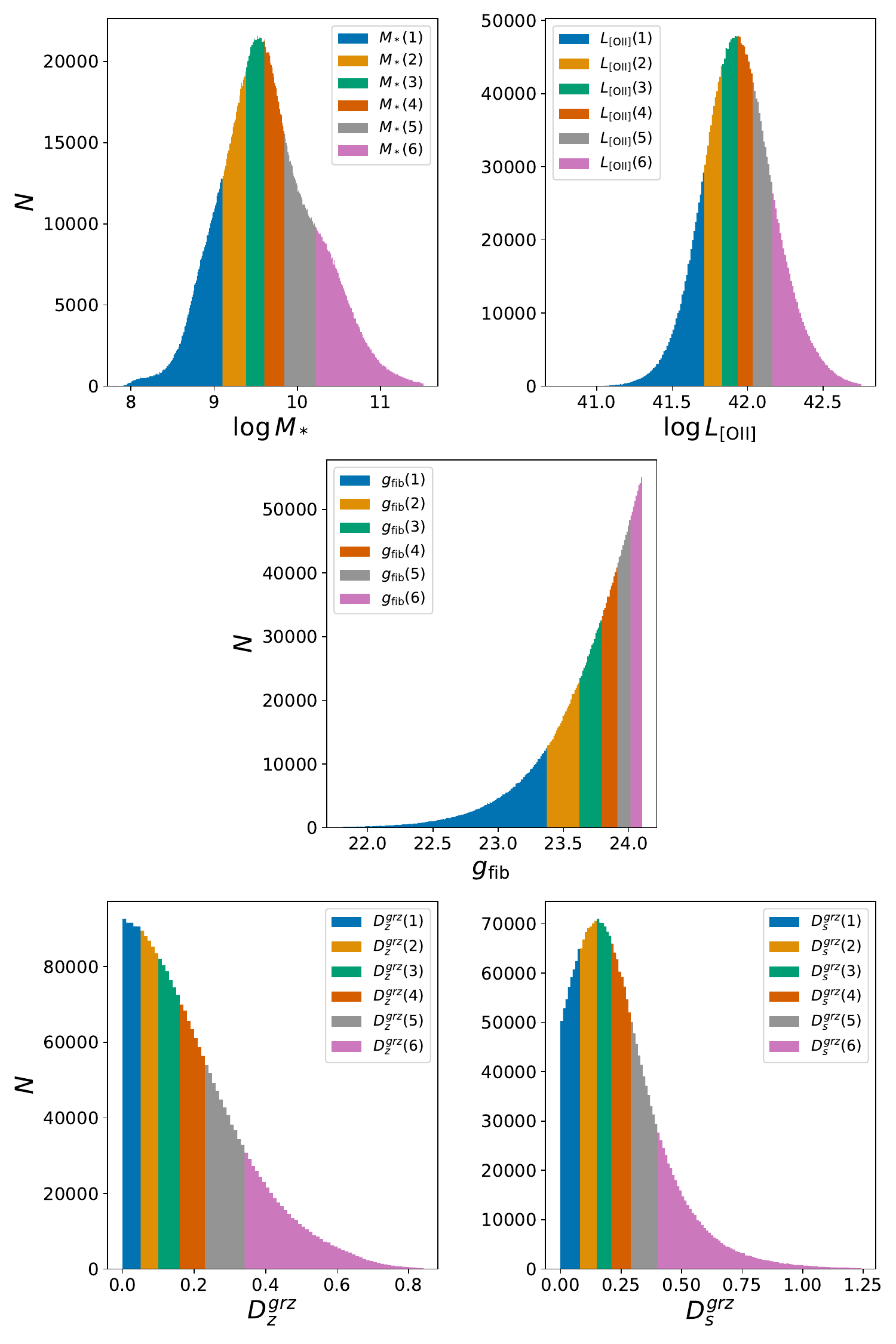}
    \caption{Subsamples defined by bounds along each galaxy property, as listed in Table~\ref{tab:samples}. All ELGs with $0.8<z<1.1$ are used when splitting on each property, meaning the sum of all subsamples in a given panel is preserved. The coloring marks distinct subsamples and matches that of all subsequent figures.}
    \label{fig:samples}
\end{center}
\end{figure*}

\begin{figure*}
\begin{center}
    \includegraphics[width=0.9\textwidth]{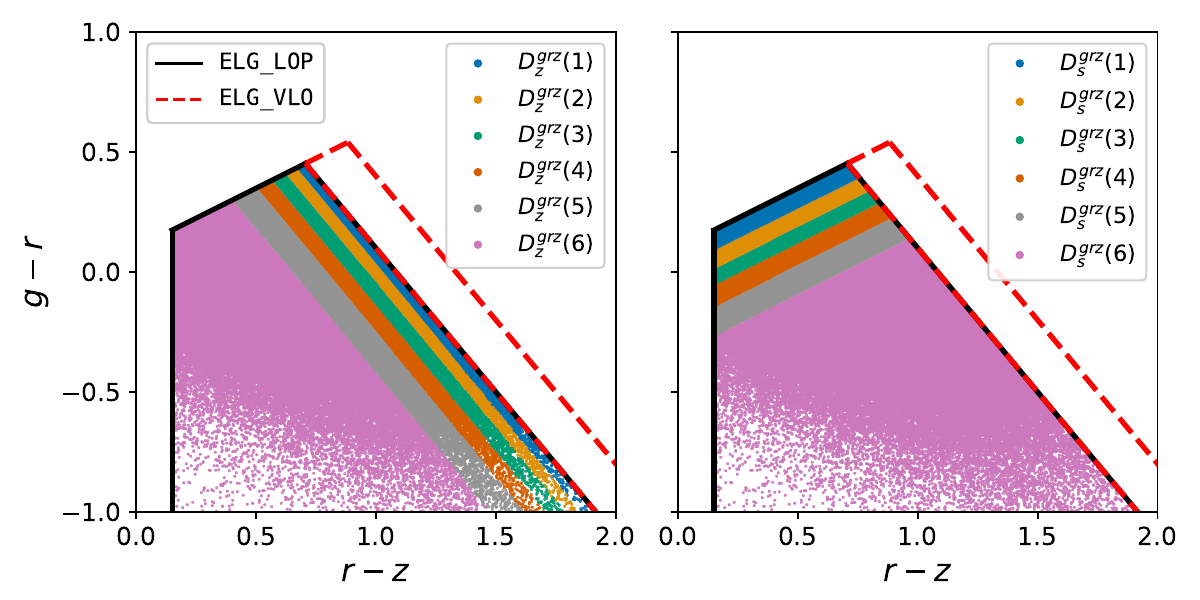}
    \caption{Confirmed ELGs with redshift $0.8<z<1.1$, with $g-r$ vs. $r-z$ boundaries defined by \dz{} and \dstar{} shown in the left and right panels, respectively.}
    \label{fig:grzpos}
\end{center}
\end{figure*}

We define subsamples according to each of the ELG properties identified in Section~\ref{subsec:properties}: \lgms{}, \lgl{}, \gfib{}, \dz{}, and \dstar{}.
For each property, we use quantile-based binning to define six subsamples containing an approximately equal number of galaxies.
Subsample definitions are given in Table~\ref{tab:samples} and illustrated in Figure~\ref{fig:samples}.
To further illustrate the meaning of \dz{} and \dstar{}, we plot the $g-r$ vs. $r-z$ distribution of the corresponding subsamples in Figure~\ref{fig:grzpos}.

\begin{table}
    \centering
    \begin{tabular}{|l|ccccc|}
    \hline
    Subsample & $\langle z \rangle$ & $\log M_{*,~\rm min}$ & $\log M_{*,~\rm max}$ & $\langle \log M_* \rangle$ & \Ngal{} \\
    \hline
    $M_*(1)$ & 0.93 & $-\infty$ & 9.10 & 8.83 & 454,969 \\
    $M_*(2)$ & 0.95 & 9.10 & 9.38 & 9.25 & 455,210 \\
    $M_*(3)$ & 0.96 & 9.38 & 9.60 & 9.49 & 462,466 \\
    $M_*(4)$ & 0.97 & 9.60 & 9.84 & 9.71 & 449,573 \\
    $M_*(5)$ & 0.96 & 9.84 & 10.22 & 10.02 & 460,605 \\
    $M_*(6)$ & 0.97 & 10.22 & $\infty$ & 10.56 & 456,607 \\
    \hline
    Subsample & $\langle z \rangle$ & $\log L_{\rm [OII],~min}$ & $\log L_{\rm [OII],~max}$ & $\langle \log L_{\rm [OII]} \rangle$ & \Ngal{} \\
    \hline
    $L_{\rm [OII]}(1)$ & 0.90 & $-\infty$ & 41.71 & 41.19 & 450,069 \\
    $L_{\rm [OII]}(2)$ & 0.93 & 41.71 & 41.83 & 41.77 & 448,501 \\
    $L_{\rm [OII]}(3)$ & 0.95 & 41.83 & 41.93 & 41.88 & 466,640 \\
    $L_{\rm [OII]}(4)$ & 0.97 & 41.93 & 42.03 & 41.98 & 457,078 \\
    $L_{\rm [OII]}(5)$ & 0.98 & 42.03 & 42.16 & 42.09 & 453,321 \\
    $L_{\rm [OII]}(6)$ & 0.99 & 42.16 & $\infty$ & 42.30 & 463,821 \\
    \hline
    Subsample & $\langle z \rangle$ & $g_{\rm fib,~min}$ & $g_{\rm fib,~max}$ & $\langle g_{\rm fib} \rangle$ & \Ngal{} \\
    \hline
    $g_{\rm fib}(1)$ & 0.95 & $-\infty$ & 23.37 & 23.01 & 462,234 \\
    $g_{\rm fib}(2)$ & 0.95 & 23.37 & 23.62 & 23.51 & 439,308 \\
    $g_{\rm fib}(3)$ & 0.96 & 23.62 & 23.79 & 23.71 & 476,443 \\
    $g_{\rm fib}(4)$ & 0.96 & 23.79 & 23.91 & 23.85 & 444,415 \\
    $g_{\rm fib}(5)$ & 0.96 & 23.91 & 24.01 & 23.96 & 448,477 \\
    $g_{\rm fib}(6)$ & 0.96 & 24.01 & $\infty$ & 24.06 & 468,553 \\
    \hline
    Subsample & $\langle z \rangle$ & $D^{grz}_{z,~\rm min}$ & $D^{grz}_{z,~\rm max}$ & $\langle D^{grz}_z \rangle$ & \Ngal{} \\
    \hline
    $D^{grz}_z(1)$ & 0.94 & 0 & 0.05 & 0.02 & 457,698 \\
    $D^{grz}_z(2)$ & 0.94 & 0.05 & 0.10 & 0.07 & 433,970 \\
    $D^{grz}_z(3)$ & 0.95 & 0.10 & 0.16 & 0.13 & 465,386 \\
    $D^{grz}_z(4)$ & 0.95 & 0.16 & 0.23 & 0.19 & 444,290 \\
    $D^{grz}_z(5)$ & 0.96 & 0.23 & 0.34 & 0.28 & 473,809 \\
    $D^{grz}_z(6)$ & 0.99 & 0.34 & $\infty$ & 0.46 & 464,277 \\
    \hline
    Subsample & $\langle z \rangle$ & $D^{grz}_{s,~\rm min}$ & $D^{grz}_{s,~\rm max}$ & $\langle D^{grz}_s \rangle$ & \Ngal{} \\
    \hline
    $D^{grz}_s(1)$ & 0.92 & 0 & 0.08 & 0.04 & 462,860 \\
    $D^{grz}_s(2)$ & 0.94 & 0.08 & 0.15 & 0.12 & 479,833 \\
    $D^{grz}_s(3)$ & 0.96 & 0.15 & 0.21 & 0.18 & 417,079 \\
    $D^{grz}_s(4)$ & 0.97 & 0.21 & 0.29 & 0.25 & 477,047 \\
    $D^{grz}_s(5)$ & 0.97 & 0.29 & 0.40 & 0.34 & 434,259 \\
    $D^{grz}_s(6)$ & 0.98 & 0.40 & $\infty$ & 0.58 & 468,352 \\
    \hline
    \end{tabular}
    \caption{Definitions of ELG subsamples with $0.8<z<1.1$. Horizontal lines separate the galaxy property used to define each group of subsamples. The leftmost column lists the shorthand name for each subsample, which we use throughout this work. The number of galaxies in each subsample is listed as \Ngal{}. Stellar masses and [OII] luminosities have units of $M_\odot$ and erg s$^{-1}$, respectively.}
    \label{tab:samples}
\end{table}

For each subsample, we use the DESI LSS catalog production code\footnote{\url{https://github.com/desihub/LSS}} to produce customized random catalogs and systematic weights necessary for accurate clustering measurements \citep{ross25}.
To generate the random catalogs, angular positions are randomly drawn from a uniform distribution across the sky and filtered to match the DESI survey geometry (e.g., footprint and masked regions around bright stars).
Redshifts (and accompanying weights; see Section~\ref{subsec:wother}) are randomly assigned from the data to match the redshift distribution.
In the case of ELGs, this step is done separately for objects in the North and South photometric regions.
By repeating this process for each subsample separately, we further customize the randoms' redshift distribution by galaxy property, thus accounting for possible variations in the $n(z)$ across subsamples.

Importantly, systematic weights are also rederived for each subsample.
As described in Section~\ref{subsec:wsys}, our method differs from that which produced the DESI-provided systematic weights in two ways.
First, we employ the same linear regression used for all other DESI tracers, while the fiducial choice for ELGs uses the \textsc{SYSNet} neural network.
Second, and of particular interest, we perform the regression on each subsample separately, allowing customized correction for variable imaging conditions as a function of galaxy property; in contrast, the fiducial weights were simultaneously derived on all ELGs in this redshift range and does not account for ELG diversity.

\begin{figure*}[ht]
    \centering
    \includegraphics[width=0.63\textwidth]{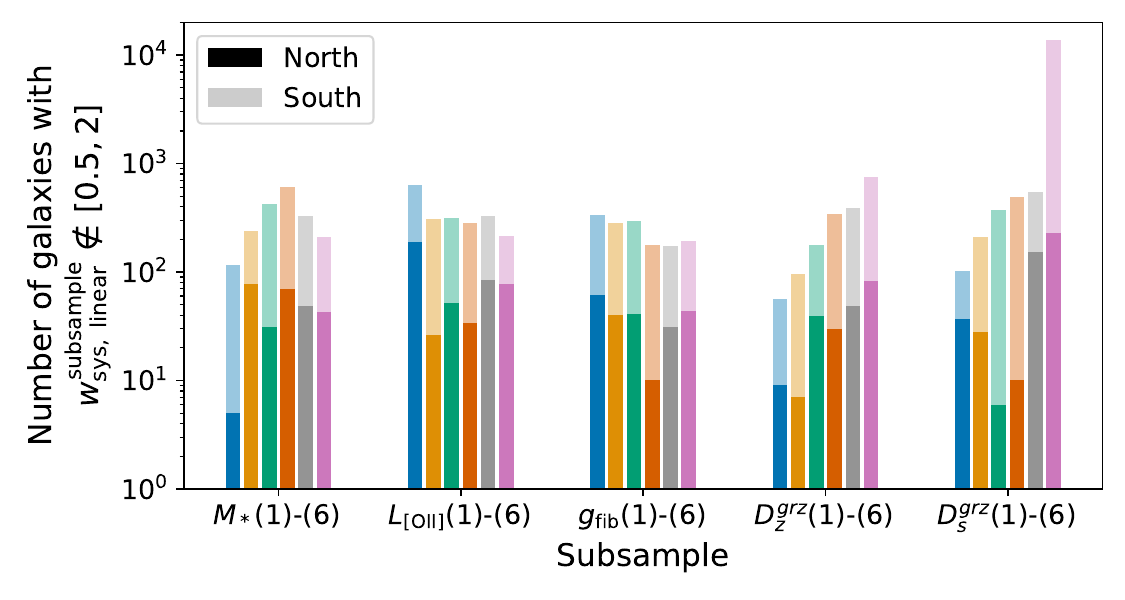}
    \hfill
    \includegraphics[width=0.36\textwidth]{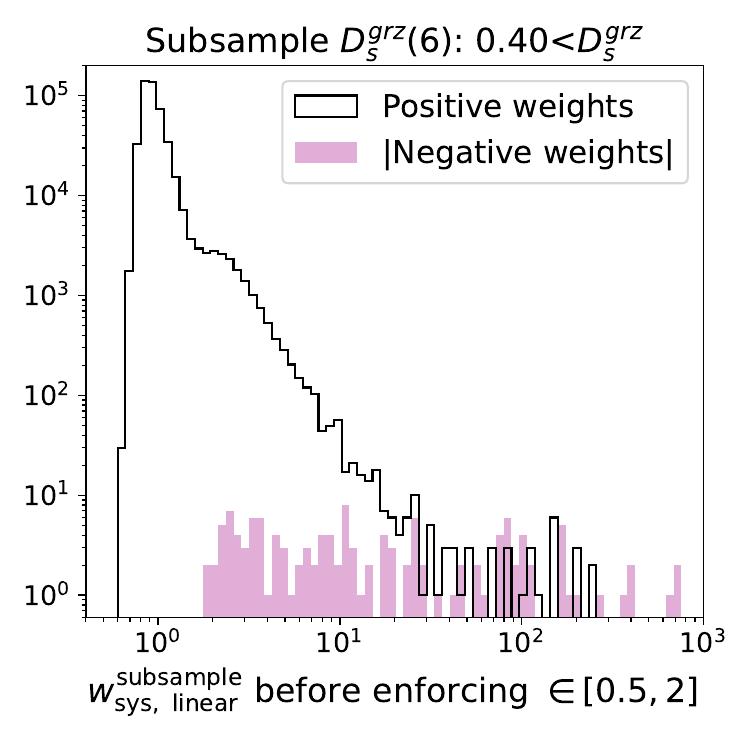}
    \caption{Inspection of systematic weights customized to each subsample. The left panel shows the number of galaxies in each subsample with weights outside $\left[ 0.5, 2\right]$. The right panel shows the distribution of new systematic weights for the subsample farthest from the stellar locus, $D^{grz}_s(6)$, before enforcing the prior. Negative weights are shown as absolute values in pink. Both panels illustrate that before enforcing the prior, the subsample farthest from the stellar locus, $D^{grz}_s(6)$, receives relatively frequent, extreme weights.}
    \label{fig:wprior}
\end{figure*}

Following \cite{rosado-marin25}, we impose a strict prior requiring that our customized systematic weights be within the range of 0.5 and 2, which limits the influence of extreme outliers that result from the linear regression.
This step is generally inconsequential, as the percentage of galaxies with weights outside this range is less than 0.17\% for all but one subsample: that farthest from the stellar locus, \dstar{}(6).
The left panel of Figure~\ref{fig:wprior} shows that this subsample has $\gtrsim$10 times more outlier galaxies than any other subsample.
The right panel of Figure~\ref{fig:wprior} shows the distribution of customized systematic weights for \dstar{}(6) before imposing the prior; despite outliers spanning approximately $-700$--250, the mean remains near unity.
Although Figure~\ref{fig:wprior} illustrates that \dstar{}(6) has relatively frequent, extreme outliers, we emphasize that such outliers constitute only $\sim$3\% of the subsample.
We therefore proceed with the prior imposed and revisit this particular subsample in Section~\ref{subsec:dstar}.

\begin{figure*}
\begin{center}
    \includegraphics[width=0.9\textwidth]{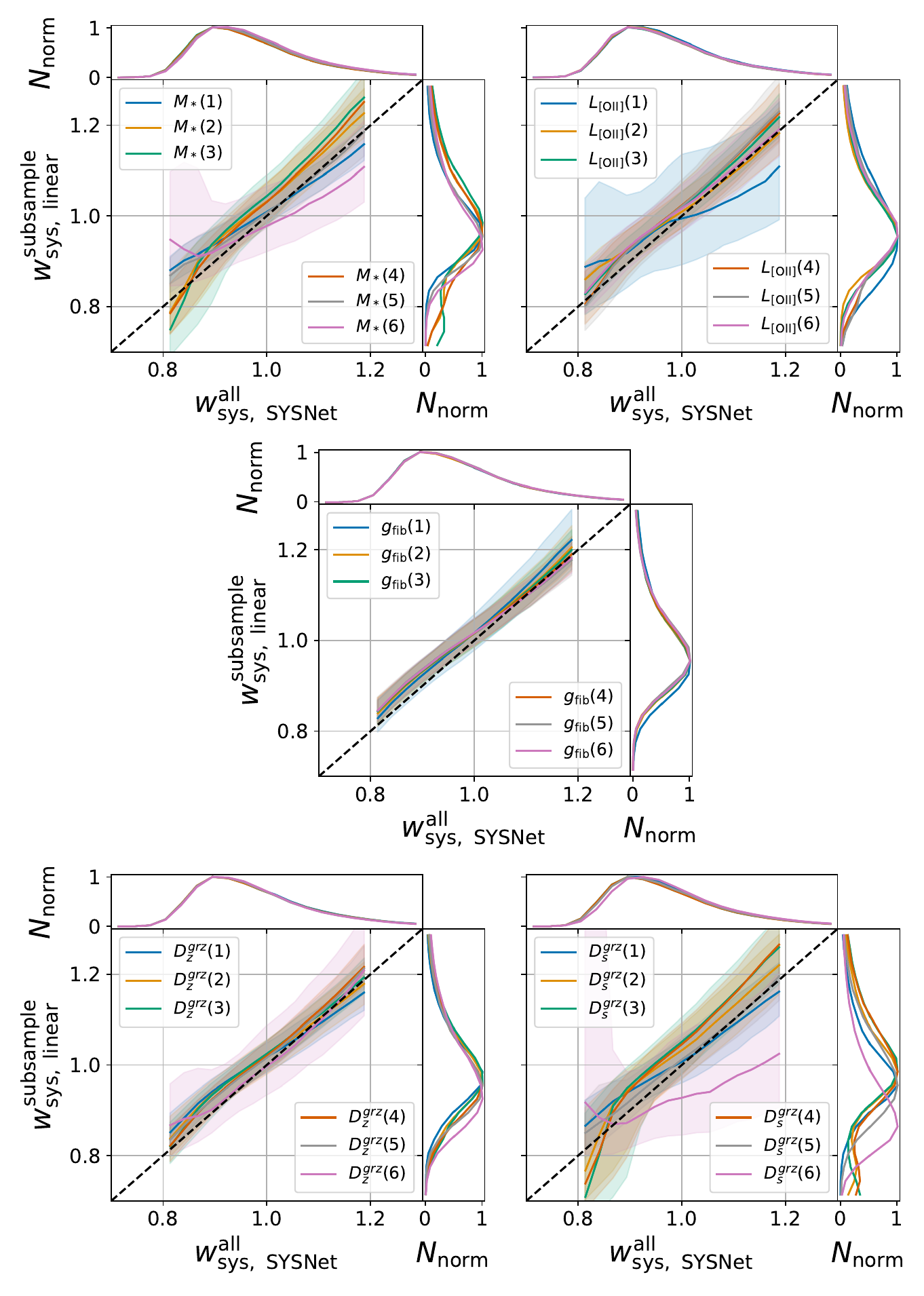}
    \caption{Comparison of our systematic weights derived via linear regression on subsamples ($w_{\rm sys,~linear}^{\rm subsample}$) to those provided in the DESI LSS catalogs, which are derived via \textsc{SYSNet} on all ELGs together ($w_{\rm sys,~SYSNet}^{\rm all}$). In each main panel, the solid line and shaded region show the median and 1$\sigma$ bounds of the distribution. Marginalized distributions---normalized to a maximum of unity---are shown along the top and right axes.}
    \label{fig:weights}
\end{center}
\end{figure*}

Figure~\ref{fig:weights} presents our property-dependent weights ($w^{\rm subsample}_{\rm sys,~linear}$) relative to their DESI-provided, fiducial counterparts ($w^{\rm all}_{\rm sys,~SYSNet}$).
Subsampling by stellar mass (top-left panel) yields appreciable change. 
In particular, the high-stellar-mass bin exhibits an asymmetric tail---as indicated by the shaded region below $w^{\rm all}_{\rm sys,~SYSNet} \lesssim0.9$---where galaxies previously receiving sub-unity weights are now boosted.
Some intermediate-stellar-mass bins also show a general increase in systematic weight.
Regarding [OII] luminosity (top-right), all but one subsample are consistent with the one-to-one relation; the weakest [OII] emitters show a broader scatter and shallower slope.
Customization to $g$-band fiber magnitude (bottom-left) and \dz{} (bottom-center) shows minimal change.
Subsampling by distance from the stellar locus, \dstar{} (bottom-right), results in the largest change.
Notably, the farthest subsample from the stellar locus, \dstar{}(6), also possesses a large tail of previously sub-unity weights that now extend to large values, as seen in the same shaded region as the high-stellar-mass subsample.
Despite this skewness, the mean weight shifts to a lower value as shown in the marginalized distribution; such a shift is unique to this subsample.

The behavior of the subsamples noted above---intermediate and high stellar mass, low [OII] luminosity, and especially those defined by \dstar{}---is largely addressed in Section~\ref{subsec:dstar}.
Before such discussion, we first present the resultant clustering measurements, which also highlight these subsamples as those with the largest deviations.

\subsection{Comparison of clustering} \label{subsec:clust_subsamps}
We measure the clustering of each subsample twice, once using the fiducial (\textsc{SYSNet}-based) systematic weights and once using our property-dependent systematic weights.
All other components of the weighting scheme, as summarized in Section~\ref{sec:weights}, are included and consistent across both measurements.

Following Eq.~(\ref{eqn:landy}), we measure the two-dimensional 2PCF as a function of pair separation perpendicular ($r_p$) and parallel ($r_\pi$) to the line of sight.
We use evenly spaced logarithmic bins in $r_p$ with a width of ${\Delta \log \left[ r_p / \left( h^{-1} \rm{Mpc} \right) \right] \sim 0.23}$.
Parallel to the line of sight, evenly spaced linear bins from $-40$ to 40~\hmpc{} (i.e., ${r_{\pi,~\rm max}=40}$~\hmpc{}) with a width of ${\Delta r_\pi \sim 2.67}$~\hmpc{} are used.
We integrate $\xi(r_p,~r_\pi)$ along the line of sight---or, in practice, sum every $i$th bin in $r_\pi$---to calculate the projected 2PCF:
\begin{equation} \label{eqn:wp}
    w_p(r_p) = \int_{-r_{\pi,~\rm max}}^{r_{\pi,~\rm max}} \xi (r_p,~r_\pi) \, {\rm d}r_\pi = \sum_i \xi (r_p,~r_{\pi,i}) \Delta r_\pi.
\end{equation}
All 2PCF measurements are performed using \textsc{Pycorr}\footnote{\url{https://github.com/cosmodesi/pycorr}}, which is DESI's wrapper of \textsc{Corrfunc} \citep{corrfunc1/2, corrfunc2/2}.
Uncertainties are estimated using 128 jackknife samples.

\begin{figure*}
\begin{center}
    \includegraphics[width=0.9\textwidth]{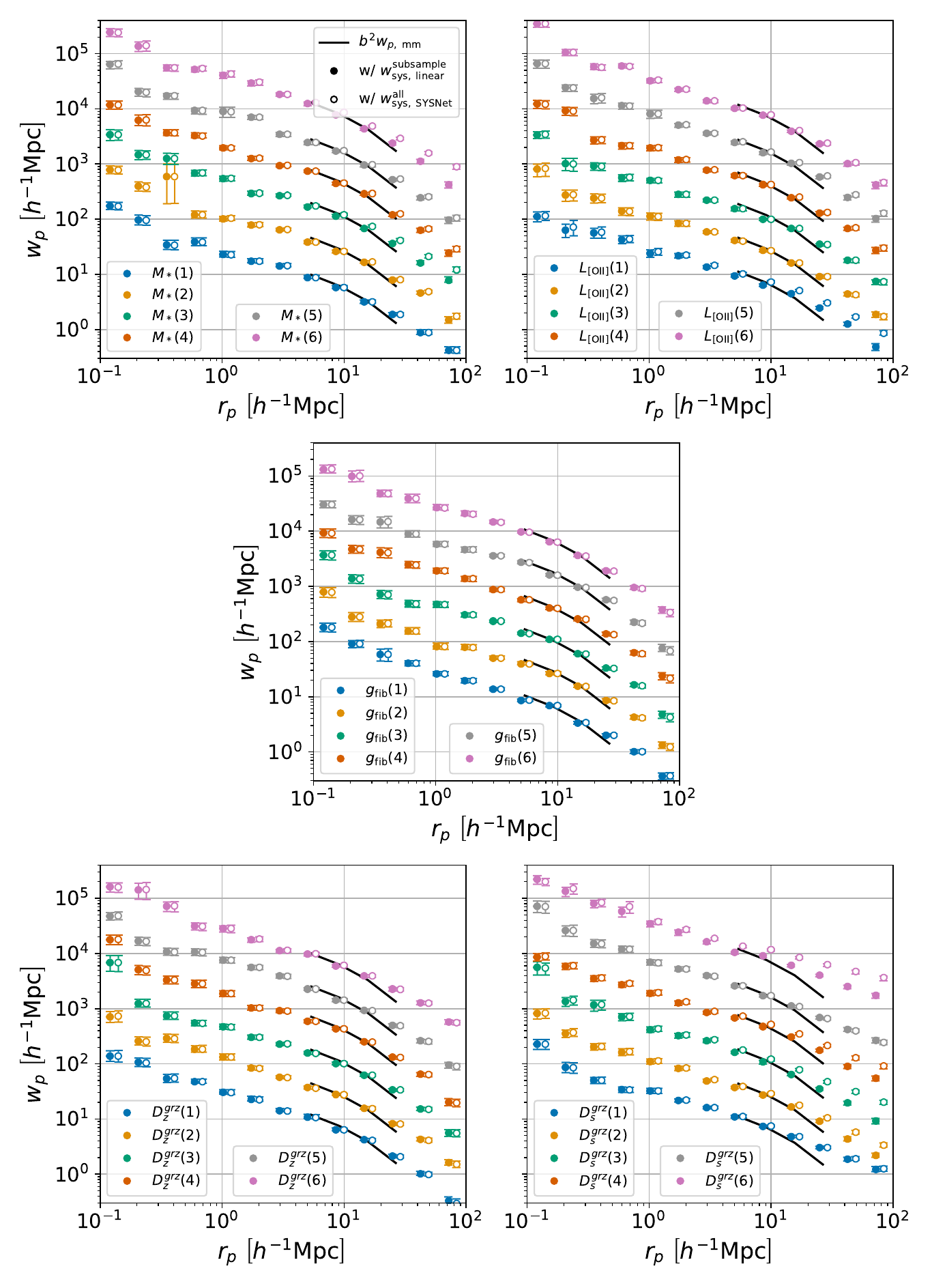}
    \caption{Projected 2PCFs using our new systematic weights ($w_{\rm sys,~linear}^{\rm subsample}$) and those using the fiducial systematic weights ($w_{\rm sys,~SYSNet}^{\rm all}$) shown as filled and open circles, respectively. For clarity, each subsample is vertically staggered by 0.6 dex, and small horizontal offsets are applied. To illustrate the expected slope of $w_p$ near $r_p\sim10$~\hmpc{}, black lines show the projected 2PCF of matter times galaxy bias ($b$) squared, fit to the filled-circle measurements on scales of $r_p=4$--$30$~\hmpc{}.}
    \label{fig:wpcompare}
\end{center}
\end{figure*}

\begin{figure*}
\begin{center}
    \includegraphics[width=0.9\textwidth]{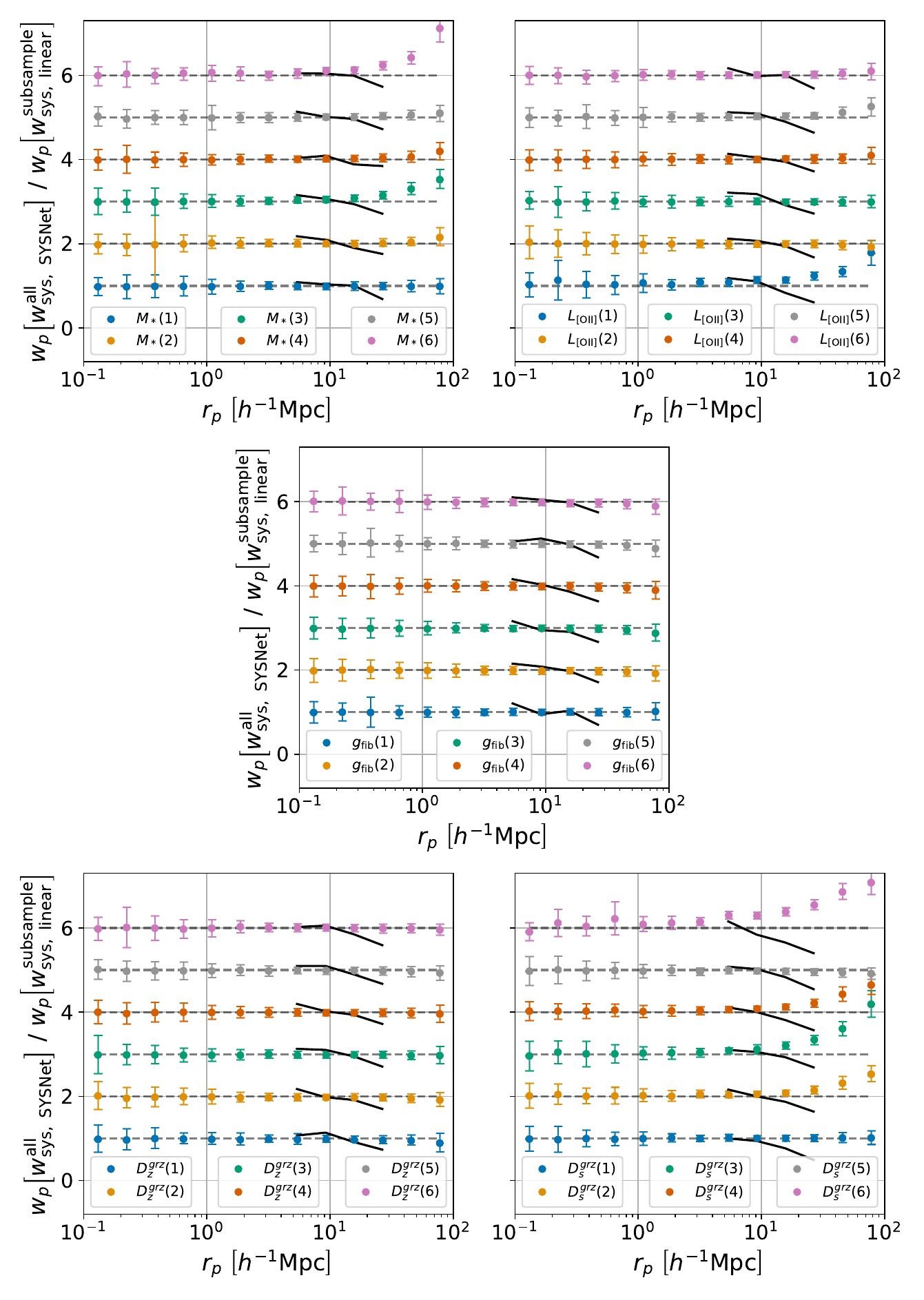}
    \caption{Projected 2PCFs using the fiducial systematic weights ($w_{\rm sys,~SYSNet}^{\rm all}$) divided by those using our new systematic weights ($w_{\rm sys,~linear}^{\rm subsample}$). Error bars show the 1$\sigma$ uncertainties, which combine those of the two measurements in quadrature. For clarity, each subsample is shifted upward by one. Dashed lines mark a ratio of unity (i.e., perfect agreement between the two measurements). Black lines show the same projected 2PCF of matter times galaxy bias squared as in Figure~\ref{fig:wpcompare}, now divided by the measurement with our new systematic weights.}
    \label{fig:wpratio}
\end{center}
\end{figure*}

Figure~\ref{fig:wpcompare} compares the projected 2PCFs using each systematic weighting scheme.
The same information is also presented in Figure~\ref{fig:wpratio}, but as the ratio of $w_p$ using the fiducial weights to those using our customized weights; doing so better illustrates the relative change in clustering.
The following conclusions can be drawn from either figure.

On scales less than $r_p \sim 10$~\hmpc{}, all subsamples have consistent clustering between the two weighting schemes.
On larger scales, the clustering of several subsamples steepen when replacing the fiducial systematic weights with our customized versions.
These subsamples are the same as those identified in Section~\ref{subsec:weight_comp}: intermediate and high stellar mass, low [OII] luminosity, and those defined by \dstar{}.
In such cases, the steepening induced by our customized weights pushes the measurement closer to an expected slope around $r_p\sim10$~\hmpc{}, suggesting that our method can further suppress spurious signal. 
To illustrate this point, we also plot the predicted $w_p$ of matter times the galaxy bias ($b$) squared (fit to $w_p$ using the customized systematic weights), the slope of which better matches those using our customized weights.
Despite this improvement, the clustering of many \dstar{} subsamples remains shallower than expected, suggesting that although the inclusion of property dependence aids mitigation, spurious signal persists.
Motivated by such behavior, we investigate \dstar{} and the corresponding subsamples in Section~\ref{subsec:dstar}, the results of which yield improved clustering measurements.

\subsection{Distance from the stellar locus and the Dark Energy Survey}\label{subsec:dstar}
Of all properties considered, subsampling by distance from the stellar locus, \dstar{}, yields (1) the most regression outliers, (2) the largest change in systematic weights, and (3) spurious signal in the projected 2PCFs.
Moreover, Figure~\ref{fig:wprior} reveals that such behavior is most prominent with ELGs farthest from the stellar locus---i.e., \dstar{}(6)---and in the South photometric region.
As shown below, much of this behavior originates from the DES footprint and can be mitigated by refining DESI's standard North-South distinction and treating DES as a third, independent region.

\begin{figure*}[ht]
    \centering
    \includegraphics[width=0.66\textwidth]{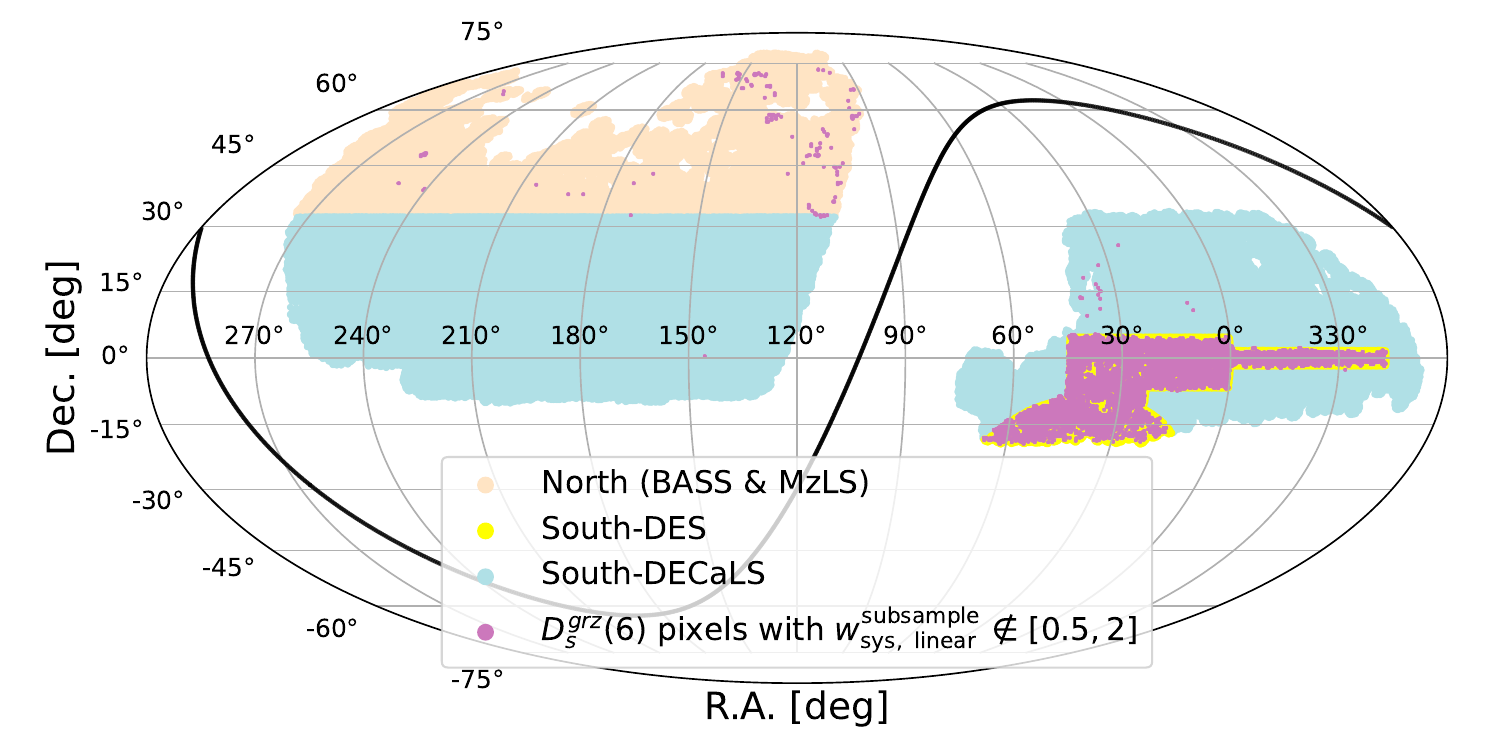}
    \hfill
    \includegraphics[width=0.33\textwidth]{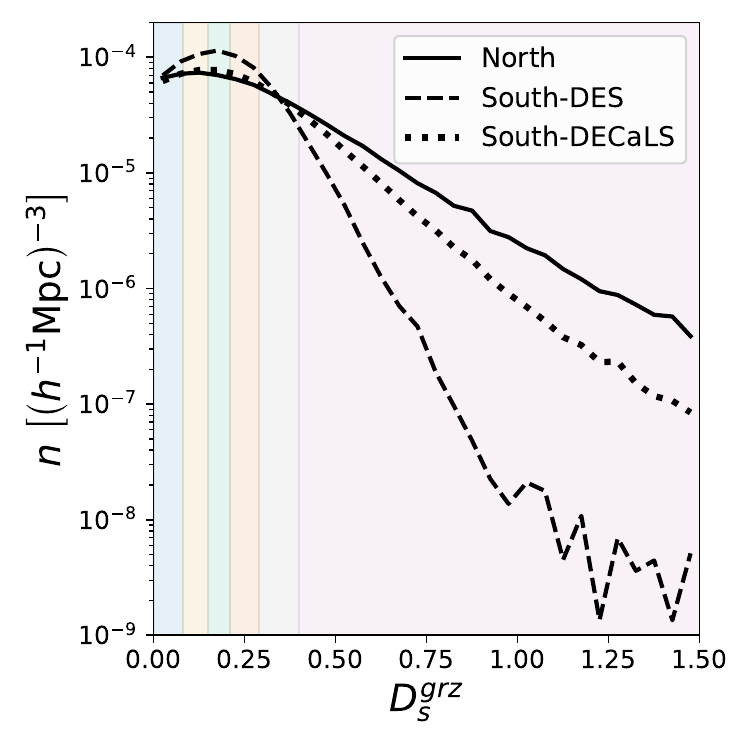}
    \caption{Separation of ELGs by source of optical imaging. The left plot shows the DR2 ELG footprint, colored by imaging source: BASS and MzLS in the North, DES in the South, and DECaLS in the South. The Galactic plane is shown in black. Overlaid as pink circles are pixels assigned a systematic weight outside the prior range when performing the regression on \dstar{}(6); majority of such pixels trace the DES footprint. The right panel shows the comoving number density within each region as a function of distance from the stellar locus. Completeness weights ($w_{\rm comp}$) are used to ensure fiber assignment does not drive the difference across regions. Background shading indicates the boundaries of the \dstar{} subsamples.}
    \label{fig:des}
\end{figure*}

Figure~\ref{fig:des} (left panel) shows the DR2 ELG footprint colored by photometric region, with the South split into South-DES and South-DECaLS.
Overlaid are pixels from the \dstar{}(6) subsample that yield systematic weights outside the prior range; these outliers clearly trace the DES footprint. 
Moreover, the right panel of Figure~\ref{fig:des} shows the ELG number density in the North, South-DES, and South-DECaLS as a function of \dstar{} and further reveals that ELGs far from the stellar locus are absent from the DES footprint.
Because DES provides the deepest photometry, these findings suggest that DESI ELGs do not truly possess large values of \dstar{}; rather, such values are only possible via measurement scatter, which preferentially occurs in the shallower regions of the Legacy Surveys.
Thus, the farthest subsample from the stellar locus, \dstar{}(6), possesses few ELGs in the DES region, resulting in an overabundance of regression outliers.
This distinction may justify separate treatment of DES during the LSS catalog production and systematic weight derivation.

\begin{figure*}
\begin{center}
    \includegraphics[width=\textwidth]{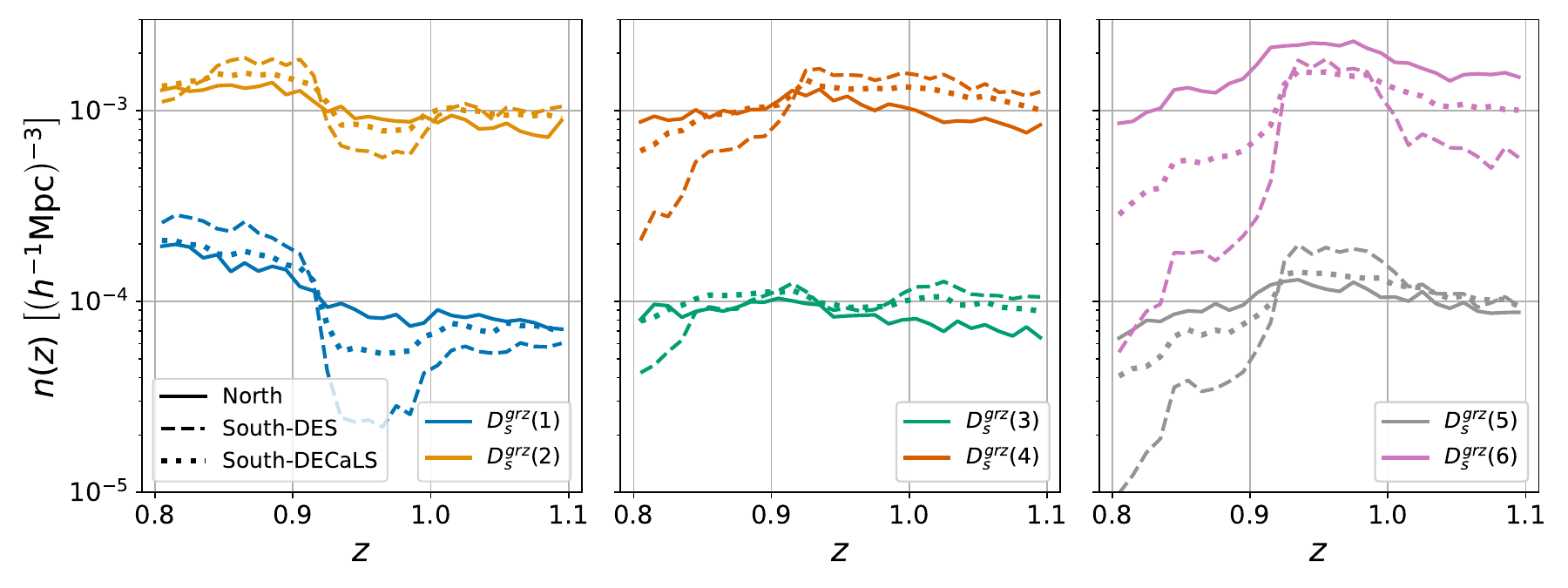}
    \caption{Number density vs. redshift for each \dstar{} subsample, split by photometric region (weighted by $w_{\rm comp}w_{\rm zfail}w_{\rm sys}$). For clarity, \dstar{}(2), \dstar{}(4), and \dstar{}(6) are shifted upward by one dex. In all cases, $n(z)$ within the DES footprint (dashed) differs appreciably from that in the remainder of the South (dotted), indicating that the redshift distribution of the corresponding random catalog should be further matched to each Southern sub-region.}
    \label{fig:nz}
\end{center}
\end{figure*}

To further motivate separate treatment of DES, Figure~\ref{fig:nz} shows the weighted $n(z)$ for each \dstar{} subsample in the North, South-DES, and South-DECaLS.
Each subsample's redshift distribution differs appreciably between the two Southern regions, meaning the standard assignment of redshifts (and weights) to the randoms---which is done separately for the North and South---is insufficient and will not capture differences between South-DES and South-DECaLS.
Notably, the redshifts in South-DES exhibit a feature at $z\sim0.95$ that transitions from a deficit to excess as \dstar{} increases.
This trend may reflect the redshift evolution of ELGs throughout the $g-r$ vs. $r-z$ plane as predicted in Figure~3 of \cite{raichoor23}, in which ELGs increase in \dstar{} up to $z \sim 1$ before returning to lower \dstar{}.
Although observable in South-DES, this feature is likely lost in the shallower regions of the Legacy Surveys via measurement scatter.

\begin{figure*}
\begin{center}
    \includegraphics[width=\textwidth]{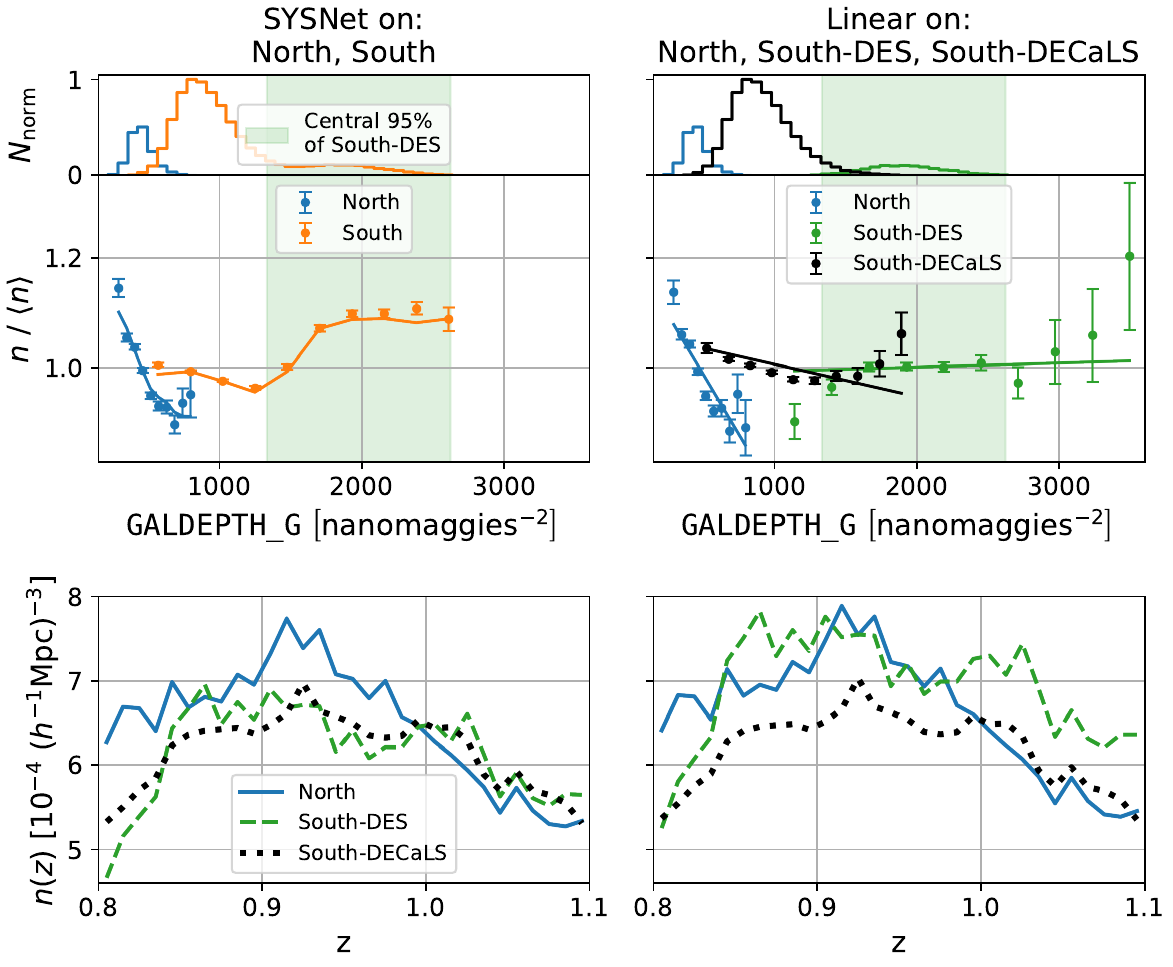}
    \caption{Comparison of fitting techniques for systematic weights. The fiducial method---\textsc{SYSNet} on the North and South---and our final strategy---linear regression on the North, South-DES, and South-DECaLS---are shown in the left and right columns, respectively. Here, we show the regression result with respect to only one of the ten imaging properties considered; upper panels show the relative ELG density (circles) vs. $g$-band depth, split into the respective regions. Model predictions are overlaid as solid curves. Marginal panels show the distribution of $g$-band depths in each region (normalized to the maximum of the South), with the central 95\% range of South-DES shaded. Lower panels show the weighted ($w_{\rm comp}w_{\rm zfail}w_{\rm sys}$) $n(z)$ in each region. While the overdensity in South-DES is suppressed by \textsc{SYSNet} (left), it is preserved by our strategy (right) and instead accounted for with customized randoms.}
    \label{fig:regress}
\end{center}
\end{figure*}

Given the differences in both abundance and redshifts between South-DES and South-DECaLS, we repeat our methodology but refine the usual North-South distinction by splitting the South into its two sub-regions.
This choice affects two key steps in our pipeline.
First, the redshifts are assigned to the randoms by separately sampling from the data in each of the three regions (with $w_{\rm reg}$ adjusted accordingly; see Section~\ref{subsec:wother}).
Second, the linear regression is also done separately for each of the three regions to produce our systematic weights.
Our final strategy is illustrated in Figure~\ref{fig:regress} and compared to the fiducial, neural-network-based approach.
In the fiducial regression (upper left panel), South-DES appears as an overdensity at high $g$-band depth, which is captured by the neural network and suppressed by the resultant weights.
In contrast, our strategy (upper right panel) produces customized randoms for South-DES that reflect the overdensity, allowing the data itself to remain unsuppressed.
The consequences are seen in the weighted $n(z)$ of each region (lower panels); \textsc{SYSNet} downweights South-DES toward the density of South-DECaLS, while our strategy preserves the density of each region.
While both strategies achieve accurate clustering measurements, our approach yields a larger (weighted) number of data pairs and maximizes sample statistics in the deepest photometric region.

\begin{figure*}
\begin{center}
    \includegraphics[width=0.9\textwidth]{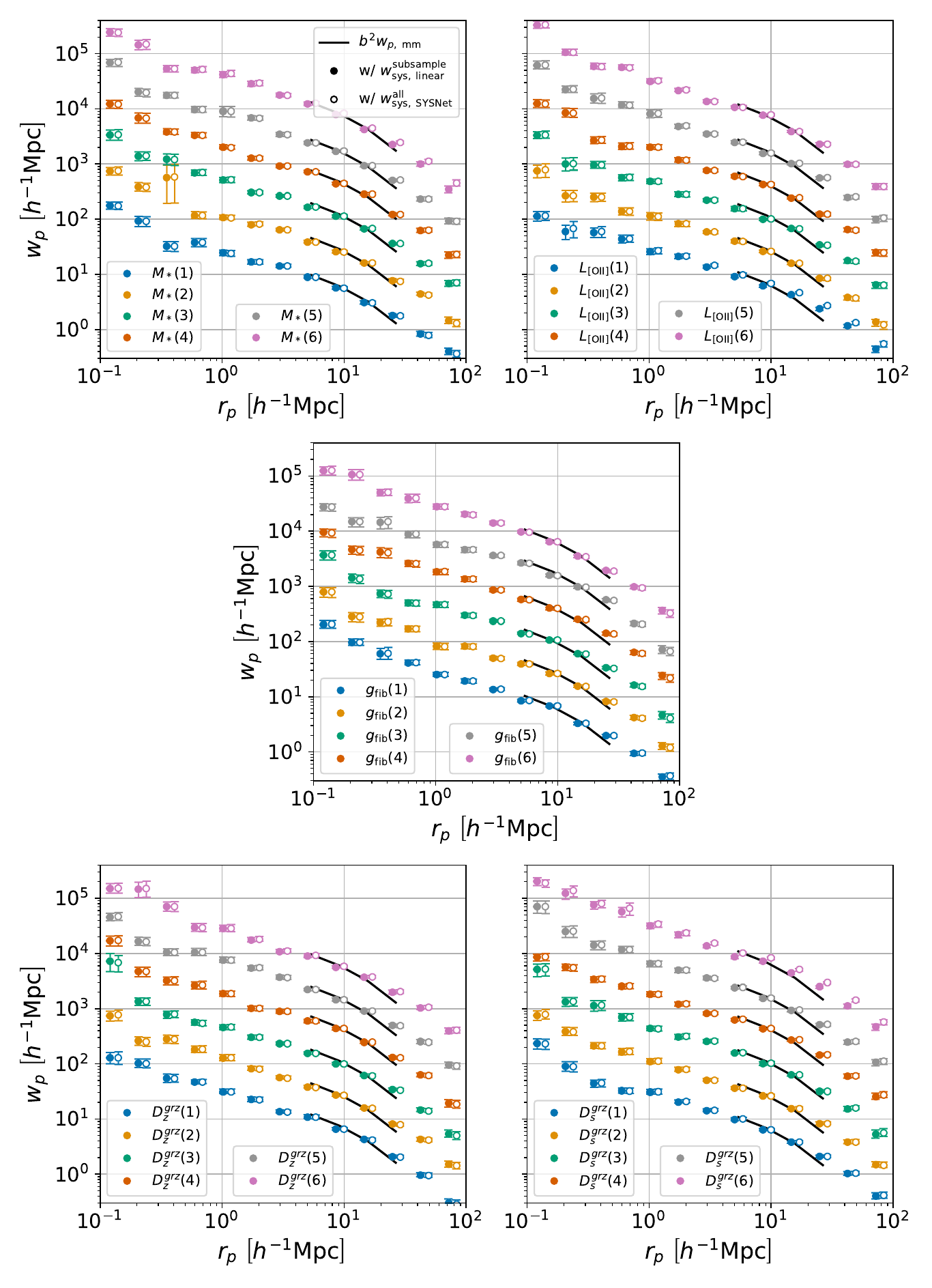}
    \caption{Same as Figure~\ref{fig:wpcompare}, but separating South-DES and South-DECaLS during the LSS catalog production and linear regression. Because the \textsc{SYSNet}-based systematic weights are not rederived, the only change with respect to the previous measurement (Figure~\ref{fig:wpcompare}) is in the redshifts of the randoms. In contrast, the measurement using our customized weights further adds a distinct linear regression for each of the three regions.}
    \label{fig:wpcompare_qso}
\end{center}
\end{figure*}

We remeasure the projected 2PCFs with the updated pipeline and show the results in Figure~\ref{fig:wpcompare_qso}\footnote{The fiducial, \textsc{SYSNet}-based weights are not rederived, thus keeping the default North-South distinction.
Therefore, when using the \textsc{SYSNet}-based weights, the only change to the pipeline is the redshifts of the randoms.}.
While the updated pipeline was motivated by the \dstar{} subsamples, new measurements are shown for all galaxy properties. 
Clear improvement is observed when comparing to the original measurements in Figure~\ref{fig:wpcompare}.
All \dstar{} subsamples possess steeper clustering, most of which now agree with the predicted slope.
Although improvement is primarily driven by the distinct treatment of DES, many of the subsamples previously identified---high stellar mass, low [OII] luminosity, and high \dstar{}---further improve when switching from the fiducial systematic weights to our property-dependent counterparts.
Therefore, we consider the combination of our customized systematic weights and DES-distinct pipeline to be the most robust, producing the most accurate projected 2PCFs.
However, we note that the projected 2PCF of \dstar{}(6) remains somewhat shallow and may still suffer from spurious signal; we consider this subsample suspect because such a large \dstar{} only occurs with shallow photometry and likely samples galaxies with seemingly unphysical colors.

\begin{figure*}
\begin{center}
    \includegraphics[width=\textwidth]{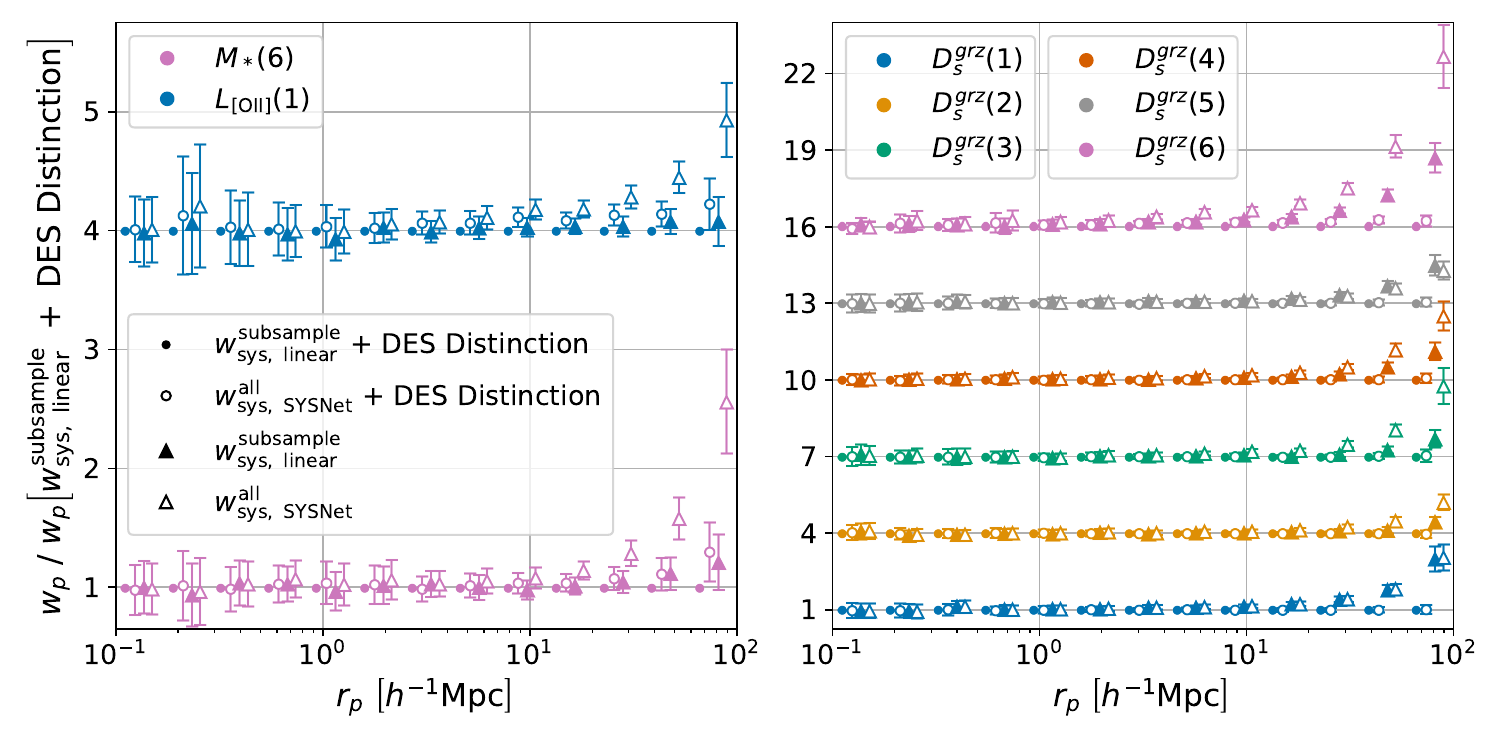}
    \caption{Projected 2PCFs using all strategies presented throughout this work, normalized by those using our most robust strategy: customized systematic weights plus distinct treatment of DES. We only show subsamples originally identified as having spurious signal: high stellar mass, low [OII] luminosity, and those defined by \dstar{}. Triangles indicate measurements following the original North-South distinction (i.e., those in Figure~\ref{fig:wpcompare}), while circles indicate those with the additional DES distinction (Figure~\ref{fig:wpcompare_qso}). Measurements using the fiducial, \textsc{SYSNet}-based weights and our customized, linear-based weights are shown as open and filled markers, respectively. Measurements for a given subsample are vertically offset and horizontally staggered for clarity.}
    \label{fig:wpsummary}
\end{center}
\end{figure*}

To summarize all improvements in our measurement, Figure~\ref{fig:wpsummary} shows the projected 2PCFs of subsamples originally identified as having spurious signal---high stellar mass, low [OII] luminosity, and those defined by \dstar{}---using each measurement strategy.
Here, measurements using the original North-South distinction (Figure~\ref{fig:wpcompare}) are shown as triangles, while those incorporating the DES distinction (Figure~\ref{fig:wpcompare_qso}) are shown as circles.
All measurements are normalized to what we consider most robust: DES distinction plus customized systematic weights per subsample.

\subsection{Impact on the monopole of all ELGs}
Thus far, we have identified multiple modifications to remove spurious signal in the projected 2PCFs of ELG \textit{subsamples}.
We now present the impact on the clustering of \textit{all} $0.8<z<1.1$ ELGs.
Doing so may reveal whether property-dependent systematic weights and separate treatment of DES play an important role in DESI's BAO and Full-Shape analysis.

For consistency with the primary DESI analysis, we measure the monopole rather than the projected 2PCF.
The monopole is obtained via the anisotropic 2PCF, $\xi(s,\mu)$, which follows the same pair-counting formula as Eq.~(\ref{eqn:landy}) but with bins of $( s,\mu )$, where $s$ is the total pair separation (i.e., $s=\sqrt{r_p^2 + r_\pi^2}$) and $\mu$ is the cosine of the angle between the pair's separation vector and line of sight.
We use evenly spaced linear bins in $s$ from 0 to 150~\hmpc{} with width $\Delta s = 4$~\hmpc{} and in $\mu$ from $-1$ to 1 with width $\Delta \mu = 0.01$.
We then decompose $\xi(s,\mu)$ into the Legendre polynomial basis to calculate the multipole moments,
\begin{equation}
    \xi_l(s) = \frac{2l+1}{2} \int_{-1}^1 \xi (s, \mu) P_l(\mu) \, {\rm d}\mu = \frac{2l+1}{2} \sum_i \xi (s, \mu_i) P_l(\mu_i) \Delta \mu,
\end{equation}
where $P_l$ is the Legendre polynomial of degree $l$, $i$ is the summation index for bins of $\mu$, and the monopole, $\xi_0(s)$, corresponds to $l=0$.

\begin{figure*}
\begin{center}
    \includegraphics[width=\textwidth]{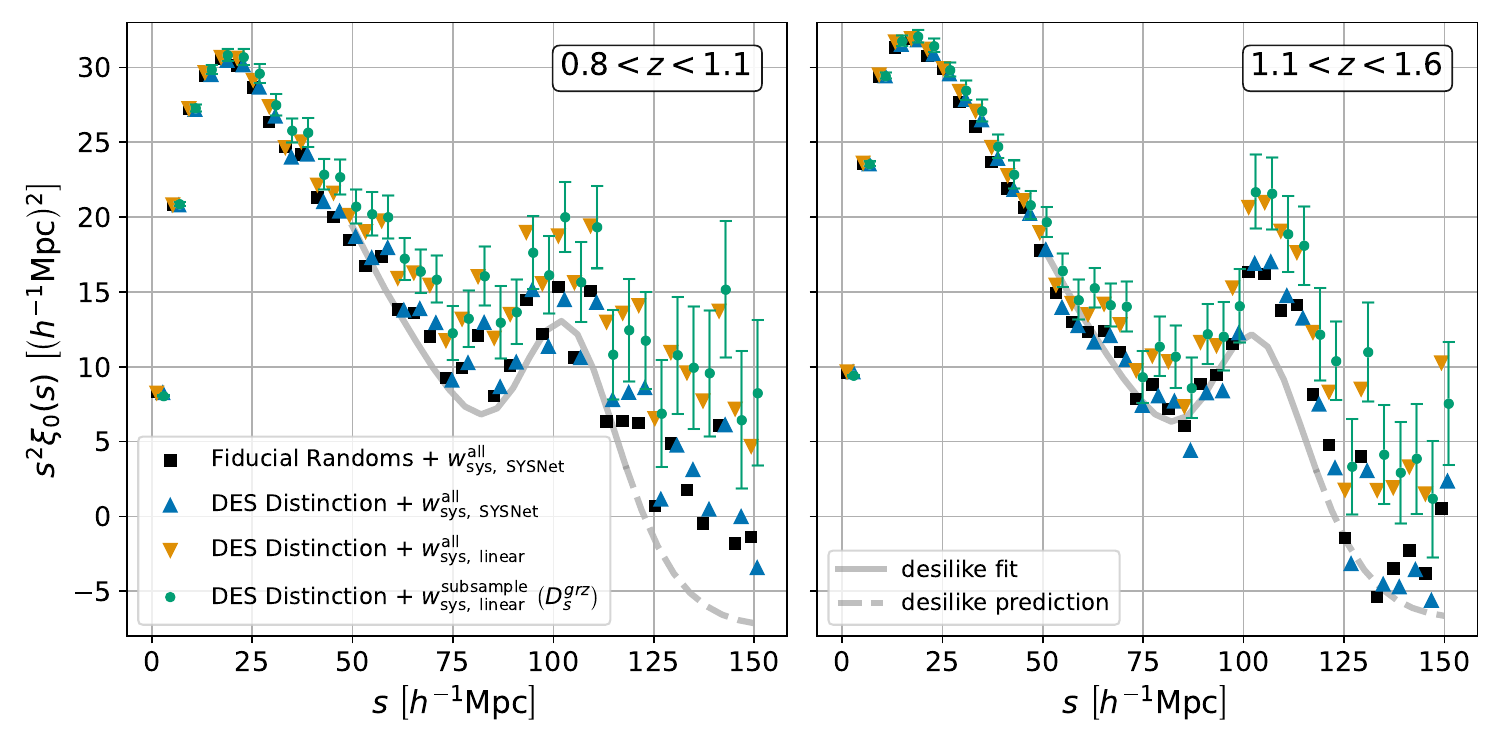}
    \caption{Monopoles of all $0.8<z<1.1$ (left) and $1.1<z<1.6$ (right) ELGs with varying measurement strategies (staggered in $s$ for illustrative purposes). Black squares are considered the fiducial DESI measurement, as the standard data, randoms, and systematic weights were used. Blue upward triangles use random catalogs whose redshift assignment in the South was further separated into South-DES and South-DECaLS. Yellow downward triangles also used DES-distinct randoms but leverage systematic weights derived via linear regression on all ELGs. Green circles show our final measurement, which uses DES-distinct randoms and systematic weights derived via linear regression on separate \dstar{} subsamples. Solid lines show the \textsc{desilike} prediction, which we fit with galaxy bias to the fiducial measurement (black squares). Dashed lines show an extension of the prediction to larger scales.}
    \label{fig:xi0}
\end{center}
\end{figure*}

We perform this measurement four times, with each iteration building upon the last.
First, we do so using the standard DESI data, randoms, and \textsc{SYSNet}-based systematic weights.
Second, we treat DES as a third, distinct region---rather than the standard North-South framework---when assigning redshifts (and corresponding weights) to the randoms.
Third, we switch from \textsc{SYSNet} to linear regression to derive the systematic weights; the regression is done on the entire sample and separated into North, South-DES, and South-DECaLS.
Fourth, we switch to our linear regression results derived on the \dstar{} subsamples; this property is chosen because it directly relates to the target selection and shows the largest variations in subsample-based clustering (e.g., Figures~\ref{fig:wpcompare},~\ref{fig:wpcompare_qso},~and~\ref{fig:wpsummary}).
The results under these four schemes are shown in Figure~\ref{fig:xi0} (left panel); we repeat our procedure with $1.1<z<1.6$ ELGs (right panel), which constitutes the DESI ``ELG2'' sample \citep{desi_dr1ii_samples}.

To summarize the difference between the fiducial result (black squares) and that using our final strategy (green circles), we compute the $\chi^2$ difference by treating the fiducial result as a model.
The resultant $\chi^2$ across 38 data points is 45.4 (53.3) at low (high) redshift, indicating the systematic difference between the two measurements is comparable (larger) than the statistical uncertainties.
As a visual guide, we also plot the prediction from the \textsc{desilike}\footnote{\url{https://github.com/cosmodesi/desilike}} package (see a brief description in Section~4.4 of \cite{desi_dr1iii_galbao}), which is fit with a galaxy bias to the fiducial measurement at $s=50$--$120$~\hmpc{} (solid line) and extended to larger scales (dashed line).
We emphasize that this prediction is largely for illustrative purposes, showing the broadband shape expected in the absence of systematic errors.
Thus, we omit any broadband correction here, highlighting that the ELG full-shape clustering measurement generally exceeds the theoretical prediction under all measurement strategies presented.

In detail, the fiducial measurement (black squares) on scales above $s\sim120$~\hmpc{} is rather consistent with the prediction at high redshift, while the low-redshift measurement suggests residual spurious signal.
Treating DES as a distinct region when assigning redshifts to randoms (blue upward triangles) yields no improvement; this result is expected because, despite some subsamples possessing different redshift distributions between South-DES and South-DECaLS, those of the combined data are similar.
Switching to linear-regression-based systematic weights---whether derived on all ELGs (yellow downward triangles) or \dstar{} subsamples (green circles)---results in excess large-scale signal.
Such an excess with respect to a neural-network-based approach is consistent with the findings of \cite{rezaie20} and is not further mitigated through the use of subsamples.
While this conclusion is true for the ELG sample \textit{as a whole}, we emphasize that property-dependent systematic weights---even those relying on linear regression---remain advantageous for \textit{subsamples}, as shown in Sections~\ref{subsec:clust_subsamps}~and~\ref{subsec:dstar}.

\section{Conclusions}\label{sec:conclusions}
In this work, we explore an alternative method for deriving the systematic weights of DESI ELG subtypes, and we do so using subsamples defined by five galaxy properties: stellar mass; [OII] line luminosity; $g$-band fiber magnitude; and two quantities describing location in the $g-r$ vs. $r-z$ plane, \dz{} and \dstar{}.
Rather than following the fiducial, neural-network-based approach, we derive systematic weights via the same multilinear regression technique used for all other DESI tracers.
We do so independently for each subsample, which adds physically-informed flexibility to the linear regression while avoiding the risks of overfitting associated with machine learning.
The consequences of this technique are explored by comparing the resultant clustering to their fiducial counterparts.
We also provide a summary of the entire DESI weighting scheme, including all individual, pairwise, and angular corrections.

While several subsamples---especially those defined by \dstar{}---benefit from customized systematic weights, we find that a more important consideration is the separate treatment of each photometric region during the DESI LSS catalog production.
Namely, the standard distinction between ELGs in the North (photometry provided by BASS and MzLS) and South (photometry provided by DES or DECaLS) is insufficient if the abundance and redshifts of a subsample vary between the South-DES and South-DECaLS regions.
In such cases, separating the South into its two sub-regions is necessary for representative redshifts in the random catalog and, given the distinct imaging properties of DES (e.g., deeper survey depth), can improve the linear regression accuracy.
ELGs that reside far from the stellar locus in $g-r$ vs. $r-z$ (i.e., large \dstar{}) are particularly sensitive to this issue, as their colors are only possible via scatter in shallow regions of the Legacy Surveys and are starkly absent in the deeper DES footprint.

We highlight three key findings.
First, the deeper imaging of DES produces a higher ELG density but far fewer galaxies with extreme $g-r$ and $r-z$ colors.
ELGs in the DES footprint also show distinct redshift behavior across the target selection plane relative to the rest of the Legacy Surveys.
Therefore, separating the South into South-DES and South-DECaLS within the DESI LSS pipeline is generally well-motivated and, in some cases, necessary.
Cuts in observed or derived galaxy properties may produce a subsample with distinct behavior in the DES region, and improper consideration of this effect can yield inaccurate clustering measurements.

Second, customizing the systematic weights to subsamples---even via linear regression---can further mitigate spurious signal; $\sim$10\% of our subsamples (high stellar mass, low [OII] luminosity, and high \dstar{}) show improved slopes in the projected 2PCF near $r_p\sim10$~\hmpc{} when custom weights are applied.
Nevertheless, the fiducial, neural-network-based weights remain the optimal mitigator of spurious signal in the clustering of the full ELG sample.

Third, photometry that places ELGs far from the stellar locus in the target selection plane is likely inaccurate, and given the stark absence in the deepest survey region, ELGs do not possess such colors in reality.
Removing such galaxies by imposing, e.g., \dstar{}$\lesssim0.4$ may produce the most consistent ELG sample (as indicated in Figures~\ref{fig:wpcompare}~and~\ref{fig:wpcompare_qso}).

These findings may be valuable to other DESI analyses.
Notably, treating DES as a separate region in the usual North-South distinction may be a justified addition for future Collaboration-wide projects.
This distinction is applicable to all target classes and is already made for the quasar sample, as the respective target selection varies with photometric survey \citep{chaussidon23}.
Other than quasars, ELGs may be the tracer most sensitive to such effects, as they are the faintest galaxy sample and approach the detection limits of the imaging surveys.
Additionally, studies which measure property-dependent clustering should use such custom systematic weights.
For example, \cite{gao23, gao24, hagen25} constrain the galaxy-halo connection of DESI ELGs by modeling the clustering of subsamples; custom systematic weights may improve clustering measurements on scales above $r_p\sim10$~\hmpc{} and affect constraints on, e.g., galaxy bias.
Finally, sample characterization is one priority for the primary DESI DR2 analyses.
Our inspection of distance from the stellar locus may provide insight into such characterization, including the ELG target selection, its dependence on photometric source, and the consequent galaxy population selected.
Careful consideration of galaxy subtypes and detailed modeling of the selection function will remain critical as larger, future DESI releases enable finer inspection and sampling of the data.

\acknowledgments
The work of T.H. and K.D. was supported in part by U.S. Department of Energy, Office of Science, Office of High Energy Physics, under Award No. DESC0009959. Z.Z. was supported by NSF grant AST-2007499.

This material is based upon work supported by the U.S. Department of Energy (DOE), Office of Science, Office of High-Energy Physics, under Contract No. DE–AC02–05CH11231, and by the National Energy Research Scientific Computing Center, a DOE Office of Science User Facility under the same contract. Additional support for DESI was provided by the U.S. National Science Foundation (NSF), Division of Astronomical Sciences under Contract No. AST-0950945 to the NSF’s National Optical-Infrared Astronomy Research Laboratory; the Science and Technology Facilities Council of the United Kingdom; the Gordon and Betty Moore Foundation; the Heising-Simons Foundation; the French Alternative Energies and Atomic Energy Commission (CEA); the National Council of Humanities, Science and Technology of Mexico (CONAHCYT); the Ministry of Science, Innovation and Universities of Spain (MICIU/AEI/10.13039/501100011033), and by the DESI Member Institutions: \url{https://www.desi.lbl.gov/collaborating-institutions}. Any opinions, findings, and conclusions or recommendations expressed in this material are those of the author(s) and do not necessarily reflect the views of the U. S. National Science Foundation, the U. S. Department of Energy, or any of the listed funding agencies.

The authors are honored to be permitted to conduct scientific research on I'oligam Du'ag (Kitt Peak), a mountain with particular significance to the Tohono O’odham Nation.

\paragraph{Data availability}
The data shown in figures are available at \url{https://doi.org/10.5281/zenodo.20600659}.
The data of Figures~\ref{fig:illustrate},~\ref{fig:samples},~\ref{fig:grzpos},~\ref{fig:weights},~and~\ref{fig:des} will be accessible once DR2 is made publicly available.

\bibliographystyle{JHEP}
\bibliography{biblio.bib}

\end{document}